\documentclass[twocolumn]{aastex631} 
\usepackage{amsmath,amstext}
\graphicspath{ {./} }

\newcommand{\JWST}{\textit{JWST}}
\newcommand{\HST}{\textit{HST}}
\usepackage[shortlabels]{enumitem}
\usepackage{needspace}

\begin{document}

\title{\JWST{} Discovery of a High-Redshift Tidal Disruption Event Candidate in COSMOS-Web}

\shortauthors{Karmen et al.}
\shorttitle{High-redshift TDE}

\correspondingauthor{Mitchell Karmen} 
\email{mkarmen1@jhu.edu}

\author[0000-0003-2495-8670]{Mitchell Karmen}\altaffiliation{NSF Graduate Research Fellow}
\affiliation{Department of Physics and Astronomy, Johns Hopkins University, 3400 N. Charles Street, Baltimore, MD 21218, USA}

\author[0000-0003-3703-5154]{Suvi Gezari}
\affiliation{Department of Astronomy, University of Maryland, College Park, MD, 20742-2421, USA}
\affiliation{Department of Physics and Astronomy, Johns Hopkins University, 3400 N. Charles Street, Baltimore, MD 21218, USA}

\author[0000-0003-3216-7190]{Erini Lambrides}\altaffiliation{NPP Fellow}
\affiliation{NASA-Goddard Space Flight Center, Code 662, Greenbelt, MD, 20771, USA}

\author[0000-0003-3596-8794]{Hollis B. Akins}
\affiliation{The University of Texas at Austin, 2515 Speedway Blvd Stop C1400, Austin, TX 78712, USA}

\author[0000-0003-3703-5154]{Colin Norman}
\affiliation{Department of Physics and Astronomy, Johns Hopkins University, 3400 N. Charles Street, Baltimore, MD 21218, USA}

\author[0000-0002-0930-6466]{Caitlin M. Casey}
\affiliation{Department of Physics, University of California, Santa Barbara, Santa Barbara, CA 93109, USA}
\affiliation{The University of Texas at Austin, 2515 Speedway Blvd Stop C1400, Austin, TX 78712, USA}
\affiliation{Cosmic Dawn Center (DAWN), Denmark}

\author[0000-0002-2361-7201]{Justin Pierel}
\altaffiliation{NASA Einstein Fellow}
\affiliation{Space Telescope Science Institute, Baltimore, MD 21218, USA}

\author[0000-0003-4263-2228]{David Coulter} 
\affiliation{Space Telescope Science Institute, Baltimore, MD 21218, USA}

\author[0000-0002-4410-5387]{Armin Rest}
\affiliation{Space Telescope Science Institute, Baltimore, MD 21218, USA}

\author[0000-0003-2238-1572]{Ori Dr. Fox}
\affiliation{Space Telescope Science Institute, Baltimore, MD 21218, USA}

\author[0009-0007-8764-9062 ]{Yukta Ajay}
\affiliation{Department of Physics and Astronomy, Johns Hopkins University, 3400 N. Charles Street, Baltimore, MD 21218, USA}

\author[0000-0001-9610-7950]{Natalie Allen}
\affiliation{Cosmic Dawn Center (DAWN), Denmark} 
\affiliation{Niels Bohr Institute, University of Copenhagen, Jagtvej 128, DK-2200, Copenhagen, Denmark}

\author[0000-0003-4761-2197]{Nicole E. Drakos}
\affiliation{Department of Physics and Astronomy, University of Hawaii, Hilo, 200 W Kawili St, Hilo, HI 96720, USA}

\author[0000-0001-7201-5066]{Seiji Fujimoto}
\altaffiliation{Hubble Fellow}\affiliation{The University of Texas at Austin, 2515 Speedway Blvd Stop C1400, Austin, TX 78712, USA}

\author[0000-0001-6395-6702]{Sebastian Gomez}
\affiliation{Center for Astrophysics \textbar{} Harvard \& Smithsonian, 60 Garden Street, Cambridge, MA 02138-1516, USA}

\author[0000-0002-0236-919X]{Ghassem Gozaliasl}
\affiliation{Department of Computer Science, Aalto University, P.O. Box 15400, FI-00076 Espoo, Finland}
\affiliation{Department of Physics, University of, P.O. Box 64, FI-00014 Helsinki, Finland}

\author[0000-0002-7303-4397]{Olivier Ilbert}
\affiliation{Aix Marseille Univ, CNRS, CNES, LAM, Marseille, France  }

\author[0000-0001-9187-3605]{Jeyhan S. Kartaltepe}
\affiliation{Laboratory for Multiwavelength Astrophysics, School of Physics and Astronomy, Rochester Institute of Technology, 84 Lomb Memorial Drive, Rochester, NY 14623, USA}

\author[0000-0002-6610-2048]{Anton M. Koekemoer}
\affiliation{Space Telescope Science Institute, Baltimore, MD 21218, USA} 

\author[0009-0003-8380-4003]{Zachary G. Lane}
\affiliation{School of Physical and Chemical Sciences --- Te Kura Mat\={u}, University of Canterbury, Private Bag 4800, Christchurch 8140, New Zealand}

\author[0000-0002-9489-7765]{Henry Joy McCracken}
\affiliation{Institut d’Astrophysique de Paris, UMR 7095, CNRS, and Sorbonne Université, 98 bis boulevard Arago, F-75014 Paris, France}

\author[0000-0003-2397-0360]{Louise Paquereau} 
\affiliation{Institut d’Astrophysique de Paris, UMR 7095, CNRS, and Sorbonne Université, 98 bis boulevard Arago, F-75014 Paris, France}

\author[0000-0002-4485-8549]{Jason Rhodes}
\affiliation{Jet Propulsion Laboratory, California Institute of Technology, 4800 Oak Grove Drive, Pasadena, CA 91001, USA}

\author[0000-0002-4271-0364]{Brant E. Robertson}
\affiliation{Department of Astronomy and Astrophysics, University of California, Santa Cruz, 1156 High Street, Santa Cruz, CA 95064, USA}

\author[0000-0002-7087-0701]{Marko Shuntov}
\affiliation{Cosmic Dawn Center (DAWN), Denmark} 
\affiliation{Niels Bohr Institute, University of Copenhagen, Jagtvej 128, DK-2200, Copenhagen, Denmark}

\author[0000-0003-2445-3891]{Matthew R. Siebert}
\affiliation{Space Telescope Science Institute, Baltimore, MD 21218, USA}

\author[0000-0003-3631-7176]{Sune Toft}
\affiliation{Cosmic Dawn Center (DAWN), Denmark} 
\affiliation{Niels Bohr Institute, University of Copenhagen, Jagtvej 128, DK-2200, Copenhagen, Denmark}

\author[0000-0002-4043-9400]{Thomas Wevers}
\affiliation{$^{}$ Astrophysics \& Space Institute, Schmidt Sciences, New York, NY 10011, USA \\}

\author[0000-0002-0632-8897]{Yossef Zenati}
\affiliation{Department of Physics and Astronomy, Johns Hopkins University, 3400 N. Charles Street, Baltimore, MD 21218, USA}
\affiliation{Space Telescope Science Institute, Baltimore, MD 21218, USA}

\begin{abstract}
The rates and properties of tidal disruption events (TDEs) provide valuable insights into their host galaxy central stellar densities and the demographics of their central supermassive black holes (SMBHs). TDEs have been observed only at low redshifts ($z \lesssim 1$), due to the difficulty in conducting deep time-domain surveys. In this work, we present the discovery of a high-redshift TDE candidate, HZTDE-1, in the COSMOS-Web survey with \JWST{}'s NIRCam, using a novel selection technique based on color and morphology. We outline a methodology for identifying high-z TDEs in deep infrared imaging surveys, leveraging their unique spectral energy distributions (SEDs) and morphologies of these transients. While focused on TDEs, this methodology could also be applied to find other UV-bright transients, such as superluminous supernovae (SLSNe). We apply this technique to COSMOS-Web in filters F115W, F150W, F277W, and F444W, and identify HZTDE-1, a transient point source relative to archival UltraVISTA infrared observations. If we assume it is a TDE, we estimate a photometric redshift of $z=5.02^{+1.32}_{-1.11}$. HZTDE-1 cannot be explained by reasonable supernova or AGN models. However, a SLSN at $z\gtrsim3$ can also plausibly explain this transient and would be the highest-redshift known SLSN. If confirmed with follow-up observations, HZTDE-1 would represent the highest-redshift TDE discovery to date, and suggest an enhancement of the TDE rate in the high-redshift universe. Our method, which can be applied to future deep surveys with \JWST{} and Roman, offers a pathway to identify TDEs at $z>4$ and probe black hole demographics at early cosmic times.
\end{abstract}


\section{Introduction} \label{sec:intro}

A tidal disruption event (TDE) occurs when a star passes close enough to a supermassive black hole (SMBH) that the tidal forces on the star overcome its self-gravity, and destroy the star \citep{Hills1975, Rees1988, Evans1989, Ulmer1999}. This process leads to a luminous flare ($\sim 10^{43}$erg/s), as approximately half of the mass of the star is accreted and the other half is ejected at high velocity \citep{Strubbe2009, Gezari2012}. This highly transient accretion event contrasts with the more long term, sustained accretion that is associated with active galactic nuclei (AGN) \citep{Begelman1984, Krolik1999, EHT2019}. Despite decades of dedicated efforts, several outstanding questions surrounding the formation and evolution of SMBHs persist. In particular, the uncertain origin of the first SMBHs has outlined the need for robustly measuring the early Universe SMBH mass distribution \citep{Volonteri2003, Begelman2006}. TDEs are one of the only phenomena, alongside AGN, that present a luminous observable very close to a SMBH. Unlike AGN, which prefer very massive SMBH hosts \citep{Kauffmann2003}, TDEs of main-sequence stars by non-rotating (Schwarzschild) black holes can only occur for lower mass SMBHs because the disrupting black hole has an upper mass limit of $M_{BH} < 10^8 M_{\odot}$ \citep{Hills1975}. Beyond this mass, the tidal radius $R_t = R_{\star}(M_{BH}/M_{\star})^{1/3}$ is within the Schwarzschild radius $R_s = 2GM/c^2$. The TDE rate's sensitivity to the black hole mass make TDEs an excellent probe of the SMBH mass function \citep{Stone2016, Kochanek2016}.  The TDE rate is also affected by the stellar mass function, so a redshift-evolving TDE rate could also be sensitive to the initial mass function (IMF) \citep{Maggorian1999}. Additionally, stellar dynamics calculations imply that galaxies with lower-mass SMBHs have higher intrinsic TDE rates \citep{Stone2016}. Therefore, TDEs are ideal probes of distant, lower-mass SMBHs and the stars which surround them.

 The primary tool used to measure the masses of high-redshift SMBHs, when dynamical information is not available, is spectroscopy of AGN \citep{Peterson1998, Kaspi2000}. Due to Malmquist bias, the most massive tail of the high redshift BH mass function is observed, leading to observations of SMBHs which are overmassive relative to their host galaxies \citep{Pacucci2024} and their cosmic epoch \citep{Juodzbalis2024}. \JWST{} observations are able to sample lower-luminosity AGN and thus lower-mass BHs \citep[e.g.][]{Li2025, Harikane2023agn}, but are prone to systematic biases in virial mass estimation methods \citep{Krolik2001, Shen2013, Bertemes2025}. TDEs offer an independent tool to study the less massive end of this mass function. The peak luminosities of TDEs \citep{Mockler2019}, their fallback rates \citep{Bandopadhyay2024}, and their late-time plateau luminosities \citep{Mummery2024b} have all been linked to their host SMBH mass. Furthermore, TDE rates can be a strong probe of the fraction of black holes in low-mass galaxies \citep{Stone2016}. TDEs are transients, which makes them significantly more difficult to observe. While thousands of AGN have been observed up to $z=11$ \citep{Maiolino2023a}, but only of order 100 TDEs have been spectroscopically confirmed at the time of writing this paper \citep{Gezari2021}, with the highest redshift TDE the jetted TDE 2022cmc at $z= 1.19$ \citep{Andreoni2022}.

Typically TDEs are discovered through wide-field, time-domain optical surveys \citep[e.g. the Zwicky Transient Facility (ZTF)][]{Graham2019, Bellm2019}. They are slow (lasting $100$\textendash$200$ days), blue (peaking in the UV) transients that do not typically cool as they fade \citep{Gezari2021}. TDEs last on scales of months, typically fading from peak in $\sim 200$~days \citep{vanVelzen2021}. The origin of the early-time UV/optical emission is debated, and still not fully understood \citep{Ryu2023, Dai2018, Roth2020}. In this work, we adopt a fully empirical, observation-driven model of TDE spectra and light curves.

At higher redshifts, TDEs are fainter and therefore their light curves are more difficult to observe far from maximum brightness. They are time-dilated and therefore evolve very slowly at higher redshifts. At $z>4$, their peak emission is also redshifted into the near-infrared which makes the both \JWST{} Near Infrared Camera (NIRCam) and Roman Space Telescope Wide-Field Instrument (WFI) ideal for observing them. This makes it a possibility to serendipitously observe an ongoing TDE in deep field \JWST{} imaging. In this work, we use the COSMOS-Web survey, the widest area deep \JWST{} survey \citep{Casey2022}, to search for potential high-redshift TDEs.  Detections of TDEs at $z>4$ could give valuable insights into SMBH formation and growth. They would indicate within which galaxies lower-mass SMBHs exist in the early universe, potentially distinguishing between SMBH seed models \citep{Volonteri2021}.

Previously, \citet{Kochanek2016} has used semi-empirical modeling of TDE rates to predict intrinsic and observed rates of TDEs as a function of redshift. They predict that, due to the observed ``downsizing" of AGN mass over redshift \citep[as modeled in][]{Shankar2009}, the intrinsic TDE rate will decrease by a factor of $5$ by $z=1$. As a result, they assert that the observed rate of high-redshift TDEs, given ground-based wide-field time domain surveys  \citep[focusing on Pan-STARRS1][]{Chambers2016}, will be even lower. Here, we extend to high redshifts by leveraging deep space-based imaging without the use of time series. We assert in a companion work \citet{Karmen_inprep} that, given the observed evolution of galaxy and SMBH properties with redshift, the TDE rate at high redshifts is likely greatly enhanced. As a result, in upcoming surveys such as the Roman Space Telescope High Latitude Time Domain Survey \citep{Rose2021} and the Rubin Observatory's Legacy Survey of Space and Time \citep[LSST][]{Ivezic2019} it may be possible to study populations of TDEs at high redshifts. \citet{Inayoshi2023} predicts the properties of high-redshift TDEs embedded within gas-rich AGN environments, in which the TDE occurrence rates are enhanced due to their compact and dense environments, compared to typical TDEs which occur in post-starburst galaxies. In this work, we complement this by a search exclusively for TDEs in galaxies not containing luminous AGN in order to utilize observed properties from low-redshift TDE searches, which exclude all AGN-hosting galaxies.

Throughout this work we adopt a Planck cosmology \citep{Planck2020}. We report optical magnitudes in the AB system. We correct observed photometry for galactic extinction using the \citet{Cardelli1989} extinction law, assuming $R_V = 3.1$ and the \citet{Schlafly2011} extinction map.

In this work, we report the discovery of a $z=5.02^{+1.32}_{-1.11}$ TDE candidate in the COSMOS-Web survey \citep{Casey2022}. In Section \ref{sec:methods}, we present a new methodology for identifying TDE candidates in deep imaging using colors and morphology. We compare to potential contaminants, and pose methods of confirming candidate TDEs. In Section \ref{sec:cosmos} we apply this methodology to the COSMOS-Web survey data, and use historic imaging of the COSMOS field to evaluate the potential variability of objects we select. We present the properties of HZTDE-1, a transient high-redshift TDE candidate in Section \ref{sec:candidate}. We compare the TDE model to potential supernova models for this object. We discuss the need for spectroscopy of this object, and its implications if validated in Section \ref{sec:discussion}. We summarize our work in Section \ref{sec:conclusion}.

\needspace{3\baselineskip}
\section{Identifying High-Redshift TDEs in Deep Surveys}
\label{sec:methods}

In this section, we model the signatures of TDEs similar to the local population, and develop a method for identifying them in wide field infrared surveys using the colors and morphologies. To first order, the UV/optical SED of a TDE is well-approximated by a black body with a typical temperature of $\sim 10^{4.3 \pm 0.15}~K$ \citep{vanVelzen2020, vanVelzen2021, Hammerstein2023, Yao2023}. We simulate TDEs with the the same temperature distribution as the latest sample of spectroscopically confirmed TDEs from ZTF \citep{Yao2023}, and simulate their volumetric density by integrating over the double-power law fit to the ZTF $g$-band luminosity function (their Equation 16).  TDEs typically do not cool as they fade, so we assume constant-temperature blackbodies for their entire evolution. Their light curves are observationally well-parameterized in \citet{vanVelzen2021} as follows: 

    \begin{align}
    L_\nu(t) &= L_{\nu_0\, \rm peak}~\frac{B_\nu(T_0)}{B_{\nu_0}(T_0)} \nonumber \\
      &\times \begin{cases} e^{-(t-t_{\rm peak})^2/2\sigma^2} & t\leq t_{\rm peak} \\ 
      e^{-(t-t_{\rm peak})/\tau} & t>t_{\rm peak}\\
      \end{cases}
      \label{eq:lightcurve} 
    \end{align} 
    
\noindent with the luminosity in the $g$-band in erg s$^{-1}$ $42.68 < \log(L_g) < 44.68$, the Gaussian rise time $0.4 < \log(\sigma) < 1.3$ days and the exponential decay time $1.2 < \log(\tau) < 2.3$ days from ZTF. When simulating TDEs, we step over redshift in shells of size $\Delta z=0.2$. At a given $z$, we sample over their temperatures and luminosities in a grid starting with the faintest TDE at a given redshift that is visible given the survey limits. We sample their luminosities in step sizes inversely proportional to their probability densities, giving more dense sampling to the higher-probability luminosities. At each redshift and luminosity, we simulate a TDE with every temperature observed in the \citet{Yao2023} ZTF TDE sample. At each given luminosity, we allow TDEs to fade as $L \propto t^{-5/3}$ until they reach the magnitude limit of the filter/survey observing them, to account for the time evolution of the TDE. 

\subsection{The Host model} 
\label{sec:hosts}

\begin{figure}
    \centering
    \includegraphics[width=1.05\linewidth]{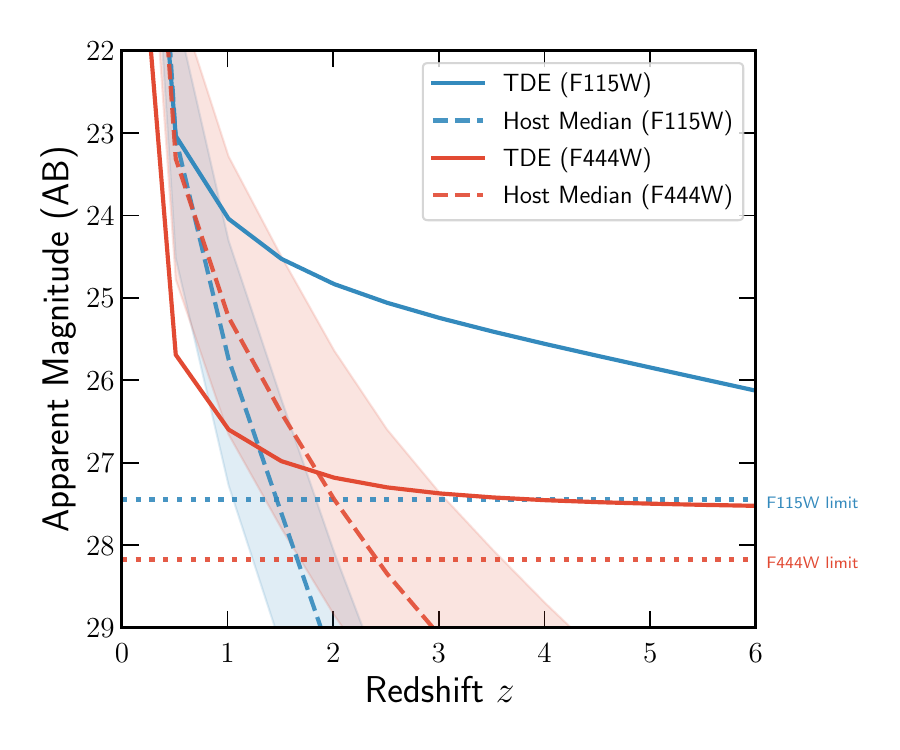}
    \caption{Apparent magnitudes of the median TDE from our simulations (solid lines) compared to the median ZTF host galaxy \citep[dashed lines ][]{Hammerstein2021}. The shaded regions represent the 5th and 95th percentile for the TDE hosts. 95\% of the hosts drop out by $z\sim2.2$ in F115W, and by $z\sim4.1$ in F444W.}
    \label{fig:host-dropout}
\end{figure}

We base our predictions of properties of TDE host galaxies on local observations of TDE hosts. It has been observed that TDE hosts are disproportionately in the green valley with compact nuclei \citep{Hammerstein2021}, and may show a preference for post-starburst galaxies \citep{French2016}. Because of this, we base our host galaxy models on the total sample of 30 ZTF TDE hosts from \citep{Hammerstein2021} to capture their observed properties. To extend the observed host spectra to their rest-frame UV, we use \texttt{Prospector} to model their spectra and perform synthetic photometry \citep{Johnson2021}. We use the parameters from the original fits of the ZTF TDE hosts in \citet{Hammerstein2021}, which are estimated by maximum a posteriori fits of Flexible Stellar Population Synthesis models \citep{Conroy2009} to broadband photometry from SDSS, Pan-STARRS, and GALEX.  When calculating properties of TDE hosts, we simulate every TDE host from ZTF to account for their full diversity of properties. We find that, due to the k-correction and the homogeneous colors of the TDE hosts, they typically are not visible beyond $z=3.5$ as seen in Figure \ref{fig:host-dropout}.

 We also use these ZTF TDE hosts as a basis for the physical sizes for typical TDE hosts. We query SDSS photometry for the TDE hosts in ZTF, and convert their angular sizes to physical sizes \citep{Blanton2017, Hammerstein2023}. We find that ZTF TDE host galaxies contain 90\% of petrosian flux in a diameter of $8 \leq D_{90} \leq 20$ kiloparsecs. This corresponds to range of angular diameters between $1.4$ and $4.7$ arcsec at $z=8$. This angular size will increase at even higher redshifts due to the decrease in angular diameter distance with increasing redshift beyond $z \sim 1.5$ \citep{Hogg1999}. As NIRCam has a resolution of 0.07 arcsec at 2 microns, all TDE host galaxies above the survey magnitude limit will be resolvable as an extended source in COSMOS-Web. At early epochs ($z\sim 7$), galaxies are expected to be intrinsically smaller \citep{Grazian2012, Kawamata2018}, such that even TDEs in unusually bright or blue hosts will not have spatially extended emission.

 Because all of the host galaxies we observe are at $z<1$, the galaxies we use are of higher mass and more luminous than those which we expect at high redshifts. High redshift galaxies are expected to be, and now observed to be, primarily lower-mass, blue, star-forming galaxies.  While there is a minority of these galaxies which are exceptionally luminous without SMBHs above the Hills mass due to extremely active star formation \citep[e.g.][]{Franco2024, Casey2024, Castellano2024}, it would be extremely difficult to detect a UV/optical TDE in one because its star formation is bright and blue, making the TDE emission a sub-dominant component of its UV SED. Therefore, we do not consider them in our analysis. However, a population of TDEs which occur primarily in star-forming galaxies have been observed to be infrared-bright with no UV/optical component to their emission \citep{Masterson2024}, making their rates relevant to wide-field surveys using MIRI. For our search in COSMOS-Web, the local TDE hosts we observe are effectively upper limits for higher-redshift host galaxy luminosity. In summary, TDEs trace lower-mass SMBHs which trace, in turn, lower-mass galaxies. These are expected to be even less massive at high redshifts, therefore intrinsically fainter, and are k-corrected into the mid-infrared. We therefore expect the vast majority of TDEs at $z>4$ to appear hostless.

\subsection{Dust extinction}
\label{sec:dust}

Interstellar dust at high redshifts is something that is hotly debated and poorly understood. To first principles, high-redshift galaxies should have very low metallically, and should therefore have less dust attenuation than galaxies today \citep{Fisher2014}. This effect is observed in local metal-poor galaxies \citep{Madden2013}, and is beginning to be seen in \JWST{} observations of high-redshift metal-poor galaxies as well \citep{Heintz2023}. Furthermore, preliminary observations of bright $z > 10$ galaxies in the first \JWST{} surveys can be explained by very low levels of dust \citep{Finkelstein2023}.

UV/optical TDEs are not typically observed in dusty galaxies \citep{French2020, Hammerstein2021}. This may be due to selection effects, or may be due to intrinsic host galaxy preferences. Our local volumetric TDE rate primarily includes TDEs which occur in local, quiescent galaxies which will decrease in number density with redshift. Combining this effect with high-redshift dust reduction, we expect that dust attenuation will have a small effect on the appearance of TDEs at high redshift. Despite this, we still include varying levels of dust in our modeling, simulating $A_V$ up to $1.2$, which is the dustiest any observed TDE host is in \citet{Hammerstein2023} fits to host photometry. Furthermore, the observation of a TDE implies the existence of a line-of-sight to the TDE emission which is not obscured by dust. Dust attenuation therefore affects the overall rates of observed TDEs more than the emission from the TDEs which are observed.

\subsection{High Redshift TDEs}
\label{sec:high_z_tde}

\begin{figure}
    \centering
    \includegraphics[width=\linewidth]{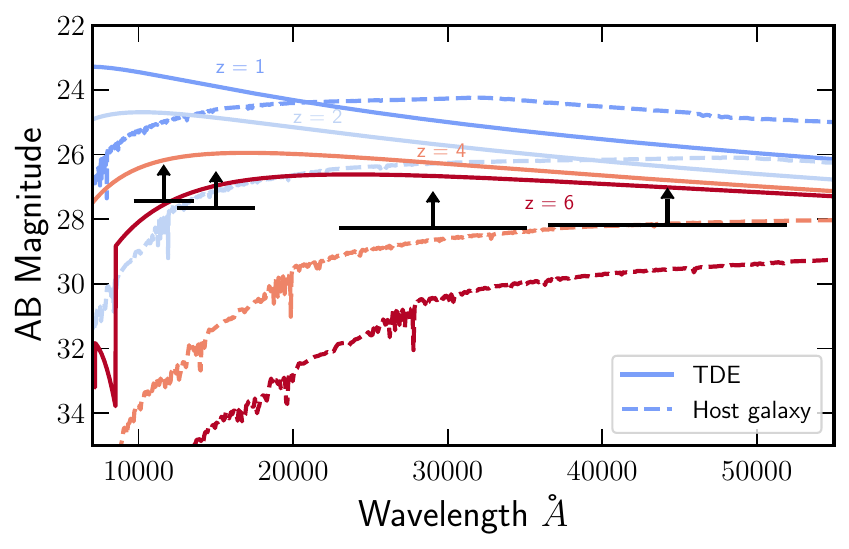}
    \caption{Example spectra of modeled $15000$~K TDEs at maximum and their host galaxies at four different redshifts: $z=1$ (dark blue), $z=2$ (light blue), $z=4$ (orange) and $z=6$ (dark red). Solid lines represent the TDE spectrum, while dashed lines show the host galaxy spectrum. The COSMOS-Web magnitude limits for each NIRCam filter are indicated in gray. At higher redshifts, host galaxies typically drop below detection thresholds, leaving TDE emission as the dominant observable component.}
    \label{fig:tde-host-model}
\end{figure}

\begin{deluxetable*}{ccccccccc}
\label{tab:fits}

\tablehead{\colhead{$C_1$ color } & \colhead{$C_2$ color} & \colhead{$a_1$} & \colhead{ $a_2$} & \colhead{$a_3$} & \colhead{Min $C_1$ color} & \colhead{Max $C_1$ color} & \colhead{Min $C_2$ color} & \colhead{Max $C_2$ color} \\ 
\colhead{} & \colhead{} & \colhead{} & \colhead{} & \colhead{} & \colhead{(mag)} & \colhead{(mag)} & \colhead{(mag)} & \colhead{(mag)}} 

\startdata
F115W - F444W & F150W - F277W & $0.0024$ & $0.4766$ & $0.0473$ & $-1.4$ & 3.5 & $-1.1$ & $1.5$ \\
F115W - F150W & F277W - F444W & $0.0242$ & $0.5215$ & $-0.6974$ & $-0.4$ & $1.3$ & $-1.0$ & $0.1$ \\
F115W - F150W & F150W - F277W & $0.0605$ & $1.3533$ & $-0.5377$ & $-0.4$ & $1.3$ & $-1.1$ & $1.5$ \\
F150W - F277W & F277W - F444W & $0.0005$ & $0.3859$ & $-0.4900$ & $-1.1$ & $1.5$ & $-1.0$ & $0.1$ \\
\enddata
\caption{Parabolas of best fit to the high-redshift TDE color-color relationship, as seen in Figure \ref{fig:color-color} anr parameterized in Eqquation \ref{eq:lines}. Each line fit is a parabola to accurately model their slight curve in color-color space. These fits are no longer valid when there is significant Lyman dropout in the F115W band.}
\end{deluxetable*}
\vspace{-1.5\baselineskip}

We find that, given our simulated TDEs and assumptions about their typical host galaxies, most high-redshift ($z>4$) TDEs will be point sources with unique colors. At $z > 4$, a typical TDE host galaxy drops in apparent magnitude below the detection limits of the COSMOS-Web survey, as seen in Figure \ref{fig:tde-host-model}. A TDE near peak brightness is visible to beyond $z \sim 10$. The SED of a high-redshift TDE is dominated by its blackbody emission, which is redshifted from the rest-frame UV to the observer-frame near-infrared NIR. Because a redshifted blackbody is another blackbody, there is a degeneracy in color-color space between temperature and redshift, meaning all idealized TDEs lie upon a nearly straight line in 2D color-color space (slightly curved due to relativistic Doppler shift and dust extinction). In our selection for high-redshift TDEs, we restrict to a region that follows the color-color relationship in any given color combination. This is shown as the three dark purple lines outlining the TDE region in Figure \ref{fig:color-color}. We also fit bounds to each individual color, which, together with the parabola fits, are presented in Table \ref{tab:fits}. We fit a parabola to the color-color relationship of TDEs in each filter combination through maximum likelihood optimization with the assumption of uniform uncertainties, which is mathematically equivalent to least squares fitting. Each curve is parameterized as:
\begin{equation}
\label{eq:lines}
    C_2 = a_1C_1^2+a_2C_1+a_3
\end{equation}

\noindent where $C_1$ and $C_2$ are two colors. The values of coefficients $a_1$, $a_2$, and $a_3$ for each pair of colors are listed in Table \ref{tab:fits}, alongside the bounds on $C_1$ and $C_2$ form our simulations. We place bounds on the redshift range we simulate, on the low end by the redshift beyond which TDEs will be point sources, and on the high end by the redshift at which Lyman absorption renders them undetectable. Additionally, we fit the minimum apparent magnitude expected for a TDE at any redshift of a given color, shown in Figure \ref{fig:color-color}, as 
\begin{equation}
    m_{F115W} = 2.15 C_{F115W-F150W} + 24.2
\end{equation}

\begin{figure*}
    \centering
    \includegraphics[width=\linewidth]{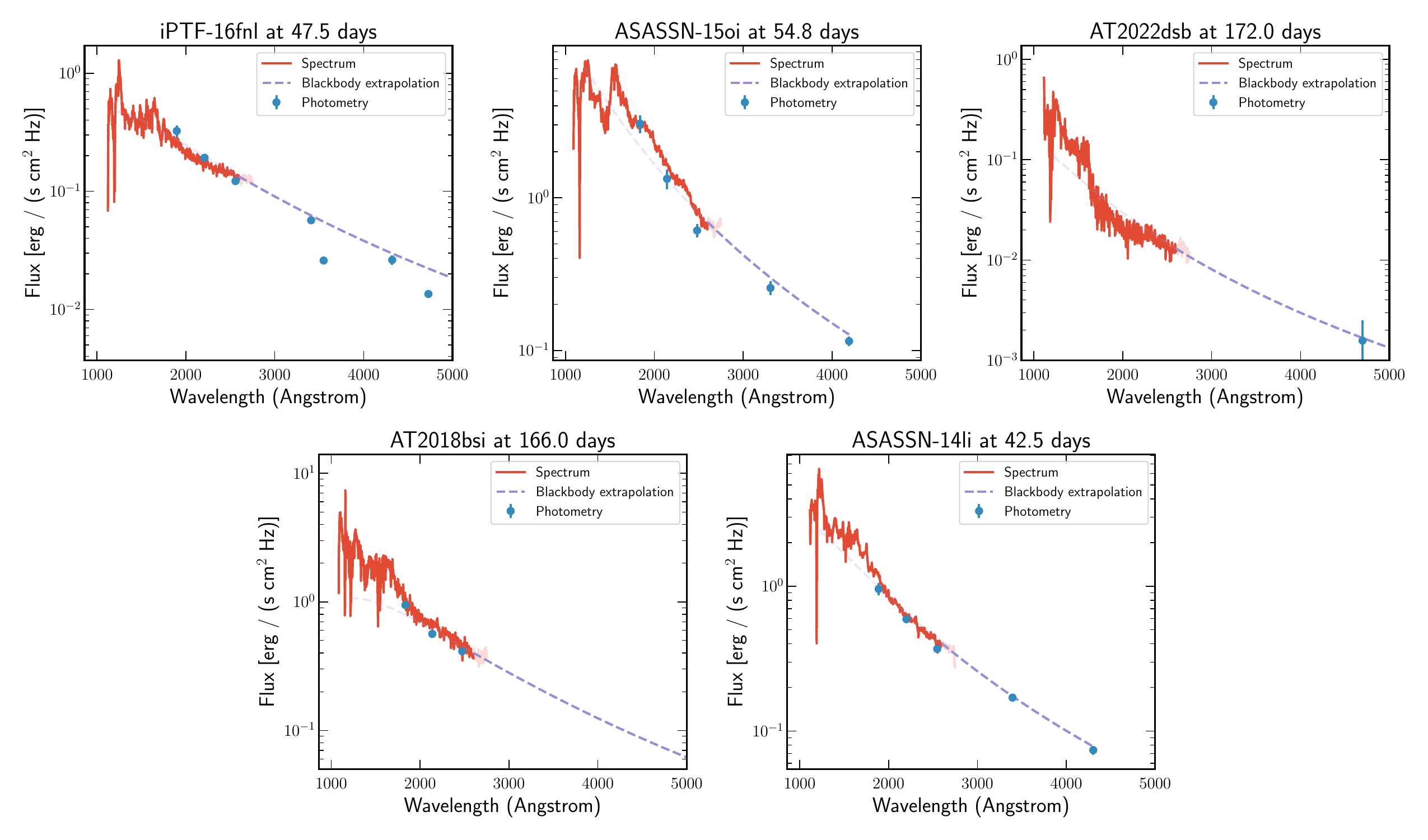}
    \caption{Example templates used to test the TDE color-magnitude cuts, shifted to the rest frame. The red line is the \HST{} spectrum, purple dotted is the blackbody used to extend the spectrum, and photometry is in blue. We switch from the \HST{} spectrum to the blackbody at rest-frame wavelength $2600 $~\AA, which is shown in the changing opacity of each line. We simulate this spectrum at higher redshifts to test our color cuts.}
    \label{fig:template-example}
\end{figure*}

To check our blackbody idealization, we take the following TDEs that have Hubble UV spectroscopy, and simulate them at high redshifts: iPTF16fnl, ASASSN-15oi, AT2022dsb, AT2018bsi, and ASASSN-14li \citep{Charalampopoulos2022, Cenko2016, Blagorodnova2019, Brown2018, Hung2019, Hoogendam2024}. The spectra used to create these templates can be accessed via \href{https://doi.org/10.17909/9m1b-2s28}{doi: 10.17909/9m1b-2s28}. We extend the UV spectrum into the optical ($>2600$~\AA) with a blackbody with the temperature fit by \citet{Mummery2024}, redshift the TDE spectra, and add in Lyman absorption shortward of rest-frame 1216\AA\ \citep{Inoue2014}. We show these templates in Figure \ref{fig:template-example}. These TDEs have colors centered on the relationship predicted, and scatter due to emission and absorption features, as well as continuum excess farther into the UV. This scatter, typically $\sim 0.2$~mag, defines the region near our idealized color-color relationships in which we expect actual TDEs to exist (the shaded region on the figure). 

While we create these selection criteria for the COSMOS-Web survey, the same process can be done for any deep, wide-field infrared survey. Of particular interest is the Roman High Latitude Wide Area Survey (HLWAS) \citep{Spergel2015}. While the High Latitude Time Domain Survey (HLTDS) is designed to identify transients \citep{Rose2021}, the HLWAS will serendipitously observe hundreds of ongoing TDEs. Using similar methodology, one could select high-redshift TDEs in the HLWAS based on their colors and morphology, and pursue follow-up to confirm them. This search serves as a first test of such a method, for application in future wide field infrared surveys. For more intermediate-redshift hostless TDEs, this method may be possible to implement in optical wavelengths with the upcoming Vera Rubin Observatory co-adds and deep drilling fields \citep{Ivezic2019}.

\subsection{Alternative TDE Models}
\label{sec:other_model} 
By extrapolating the blackbody NUV/optical spectrum of a TDE to the UV, we predict the spectrum of TDEs at rest-frame wavelengths at which their spectra are relatively poorly constrained, between $900 $~\AA\ and $1500 $~\AA.  Furthermore, the presence of Bowen lines in TDE spectra \citep[e.g.][]{Leloudas2019} implies that the flux predicted by a blackbody at $228 $~\AA\ is $\sim 5$ orders of magnitude less than what is needed in order to ionize He II and cause Bowen fluorescence.  In the intermediate wavelength that we observe, between $900 $~\AA\ and $1500 $~\AA, a blackbody may also under-predict the TDE flux.

To account for this potential discrepancy, we also consider a TDE spectrum with enough UV flux in order to trigger Bowen lines.  For this test, we consider the \citet{Dai2018} model, which uses 3D GRRMHD simulations to simulate a super-Eddington accretion disk, and Monte Carlo radiative transfer to simulate reprocessing; we do not claim that this model is correct but treat it as an example of a spectrum which is extremely luminous in the UV. We select emission from a viewing angle between $22.5 -\:45 ^\circ$ from a simulated spectrum which matches a blackbody spectrum of a few $\times 10^4$~K in the optical. This model is not used for any rate calculations or final conclusions, but is a comparison point for TDE colors and identification methodology. 

We find that the \citet{Dai2018} model falls well within our color-color and color-magnitude bounds, typically within $0.1$~mag of the central TDE color-color lines that we fit. While it would skew our fit to that of a hotter or lower-redshift TDE, spectroscopic follow-up would confirm our TDE redshift regardless of the underlying continuum model.

\subsection{Contaminants}
\label{sec:contaminants}

\begin{figure*}

    \includegraphics[width=\textwidth]{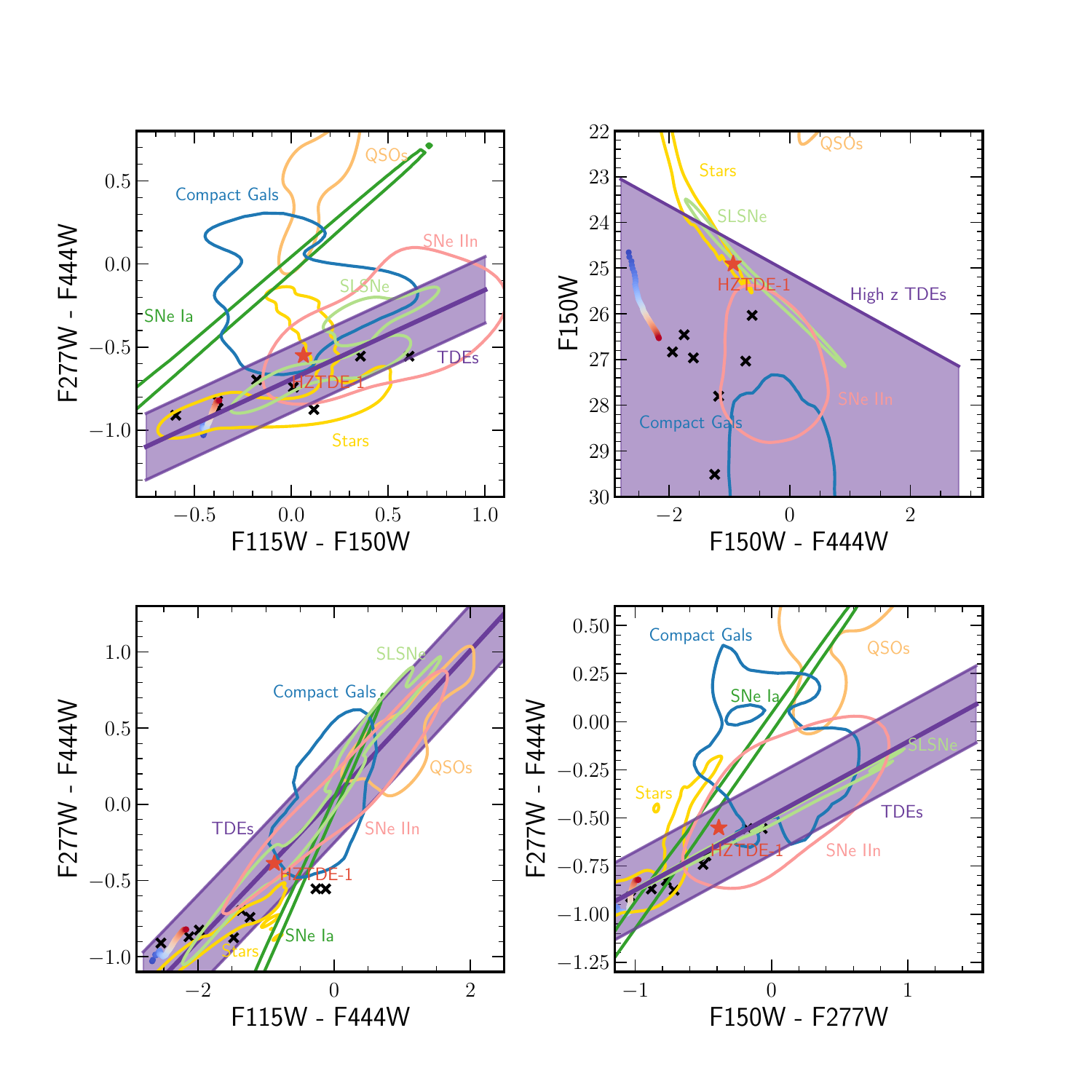}
    \caption{Color-color and color-magnitude plots of high redshift ($z>4$) TDEs compared to potential point source contaminants, modeled as described in Section \ref{sec:contaminants}. The purple central line is the parabola fit (Equation \ref{eq:lines} and Table \ref{tab:fits}) to the TDE color-color relationship, and the region surrounding it is the $0.2$~mag of scatter seen in the TDE template colors from section \ref{sec:high_z_tde}. The unshaded regions contain 95\% of contaminants of a given type. The red star is HZTDE-1, the high-redshift TDE candidate we identify in Section \ref{sec:cosmos}. The black x's are the synthetic photometry of our TDE templates at $z=5$. HZTDE-1 is notably brighter than these templates because the majority of the template UV spectra are taken weeks to months after maximum brightness. The \citep{Dai2018} model is represented in the red-blue line, with the bluest end representing $z=2$, and the reddest representing $z=8$.}
    \label{fig:color-color}
\end{figure*}

We model any point sources which may have similar colors to high-redshift TDEs to verify that high-redshift TDEs lie in a unique region of color-color and color-magnitude space. This varies dramatically by survey, but we demonstrate these color properties for COSMOS-Web, as the majority of future wide-field deep surveys will be in the near-infrared. We show examples of these color-color and color-magnitude relationships in Figure \ref{fig:color-color}. The Roman HLTDS and HLWAS may in fact be easier to identify high-redshift sources in, because optical imaging from the surveys means that a Lyman break will be observed to verify the redshift of a source. This breaks multiple degeneracies we see in the \JWST{} NIRCam infrared colors. The contaminants we model are as follows:

    \textit{Lower redshift supernovae}: SNe Ia, SNe IIP and SNe IIn at $z>2$ may be point sources in \JWST{} data. Already, supernovae with host redshifts from 0.6461 to 3.6 have been discovered using NIRCam imaging in the \JWST{} Advanced Deep Extragalactic Survey (JADES) \citep{Pierel2024, Siebert2024, DeCoursey2024, Coulter2025}. We specifically model SNe which are point sources, so we do not consider contribution in their colors from host galaxy light. Our SN Ia models are SALT3 models which extend into the NIR \citep{Pierel2022}, with parameter sweeps over $-2 \leq x_1 \leq 2$, $-0.3 \leq c_1 \leq 0.3$, and $z \leq 5$, simulated at phases $-20 \leq \textrm{p}\leq 50$ days. We predict that no SNe Ia will be detected beyond $z > 5$ in COSMOS-Web. To model SNe II, we use every SNANA model \citep{Kessler2009} that has been implemented in \texttt{sncosmo} to represent the diversity of observed SNe II at their peak luminosities. The SNANA models have only an amplitude parameters, so we sweep over a range of luminosities for each template. We set bounds on each supernova subtype's luminosity distribution based on observed samples, $-16 < M_b < -19$ for SNe Ia \citep{Phillips1993}, $-16 \leq M_V \leq -19.5$ for SNe Ib/c \citep{Drout2011}, $-16 \leq M_b \leq -18.5$ for SNe IIP \citep{Sanders2015}, $-13\leq M_g \leq -22$ for SNe IIn \citep{Nyholm2020}, and $-20 \leq M_g \leq -22$ for SLSNe \citep{Gomez2024}.

    We find that SNe Ia are sufficiently bluer than TDEs at lower redshifts where their apparent magnitudes match, and are too faint at higher redshifts when their colors match, such that there is no confusion at any redshift as seen in Figure \ref{fig:color-color}. SNe Ib/c can indeed match the colors and magnitudes of high-redshift TDEs at redshifts $z<1$, however their hosts are expected to be detected in COSMOS-Web and previous deep optical imaging of the COSMOS field. SNe II are even bluer, and at $z\sim2$ they can pass as $z>4$ TDEs, but would need rest-frame $b$-band absolute magnitudes $\lesssim-20$, which is not observed in SNe IIP. SNe IIn can become this luminous at maximum brightness, serving as our main contaminant if their hosts are not detected at $z\sim 2$. We show SNe IIn at $1.5 < z < 2.5$ in Figure \ref{fig:color-color}.

    \textit{Superluminous supernovae (SLSNe)}: SLSNe, alongside TDEs and pair-instability supernovae, will be among the few transients which are visible up to very high redshifts. We model them using the templates constructed from the Dark Energy Survey (DES) SLSNe, which have a blackbody continuum with UV absorption features superimposed \citep{Angus2019}. Because their absorption strengths evolve and the SLSNe cool after their explosion, we simulate them up from $-10$ to $+35$ rest-frame days from peak with early-time temperatures between $5000$~K and $18000$~K. We use \texttt{sncosmo} to model these and perform synthetic photometry \citep{sncosmo}. Again, we do not consider contribution from the host galaxy.

    SLSNe are persistent contaminants given the low-mass nature of their typical hosts, and their colors and luminosities which significantly overlap those of TDEs \citep{Schulze2018}. A candidate TDE of a given high redshift could be a SLSN of comparable redshift. Time domain information, or spectroscopy is needed to distinguish a candidate TDE from a SLSN: SLSNe typically cool as they fade, whereas TDEs remain constant in color. A fading source with no color changes in two to three total observations over rest-frame months (observer-frame $6$\textendash$8$ months between observations) is an extremely strong TDE candidate. Finally, spectroscopy can identify typical TDE emission lines, such as broad H-alpha and He II, which can indicate a TDE \citep{Gezari2021}. However, a class of spectral ``featureless" TDEs also exists, which would not have any strong emission line features \citep{Hammerstein2023}. Therefore, a combination of time-domain and spectroscopic monitoring is necessary for this rare case.

    \textit{Quasars/AGN}: We model both Type 1 and Type 2 AGN for comparison. We use the following SWIRE Type 2 AGN templates: QSO2, Torus and Sey2 \citep{Pollette2007}. We use the templates as they are integrated into the \texttt{pysynphot} Python package \citep{pysynphot2013}, and renormalize them to the $M_{UV}=-19.5$~mag.  This is a typical absolute magnitude for a high-redshift obscured AGN given the JADES narrow-line AGN sample \citep{Scholtz2023}. We find, however, that our selection is relatively insensitive to the luminosity function of Type 2 AGN, and depends on their SED shape alone.  Our Type 1 AGN model is a composite quasar template which combines the SDSS quasar spectral template in \citep{VandenBerk2001} with NIR quasar spectral template constructed from IRTF spectra in \citet{Glikman2006}. We calibrate the absolute magnitude of the model in the $r$-band to the high-redshift ($z=7$) QSO luminosity function in \citet{Matsuoka2023}, with a break magnitude of $-25.6$~mag. We redshift this quasar model at redshifts up to z=8, just beyond the highest-redshift confirmed quasar, J0313–1806 \citep{Wang2021_qso_z}.

    Type 2 AGN are universally too red to be contaminants in our TDE sample. A $z>8$ type 1 AGN can be similarly blue and have comparable apparent magnitude as a $z\sim 5$ TDE, but the power-law slope of its spectrum can be clearly distinguished from a blackbody given four photometric filters. Neither serves as a major contaminant in our sample.

    \textit{Stars}: We implement the PHOENIX stellar templates for stars with $0 < \log{g} < 5.5$, $2000 < T_{\textrm{eff}} < 24000$ and $-2 < \textrm{Fe}/\textrm{H} < 1$ \citep{Husser2013}. We use \texttt{pysynphot} both to generate their spectra and to perform synthetic photometry in the NIRCam filters.  We later use the same PHOENIX models to carefully simulate cool dwarf stars, such as M and L dwarfs. We simulate cool dwarfs to Milky Way halo distances, up to $20$~kpc, although distances $>5$~kpc are unlikely due to the scale height of the Milky Way, or until they are too faint for detection in COSMOS-Web.

    Ultracool dwarf stars with temperatures $2000 \leq T_{BB} \leq 3200$~K can masquerade as high redshift TDEs because they have predominantly thermal cool continuum emission which resembles redshifted hot thermal emission; e.g. a 2400~K local blackbody has the same spectral shape as a $z\sim6$ 15000~K blackbody. While the dwarf star spectral templates we have are slightly flatter than blackbodies, they become a contaminant when searching for higher-redshift ($z>6$), cooler ($< 20000$~K) TDEs. To distinguish these, we need to detect either time evolution, or a Lyman break to verify a candidate as an extragalactic transient. This is most prevalent when selecting bright ($>25$~mag) sources, but a halo star could hypothetically mimic a faint TDE candidate.

     \textit{Brown dwarfs}: We use the Sonora Bobcat models \citep{Marley2021}. These are both rare, and they do not match colors of TDEs so this comparison is just for verification of their difference. They are omitted from Figure \ref{fig:color-color} for visual clarity, because they lie far outside the axis limits.

     \textit{Point-like galaxies}: A small fraction of very-high-redshift galaxies observed in \JWST{} are compact enough to render them point-like.  We model these using the JAGUAR \JWST{} Extragalactic Mock Catalog \citep{Williams2018}. This catalog models the evolving luminosity functions, spectroscopic properties, and morphologies of galaxies up to $z\sim 15$, and can reproduce observations up to $z \sim 10$. We specifically use simulations which use templates of expected $5 < z < 15$ galaxies.  We select galaxies which will be point sources by selecting those which have an observed angular size equal to or less than the FWHM of the F115W PSF, $0.07$~arcsec. We use the pre-synthesized photometry from these catalogs for a wide range of galaxy types and redshifts, up to $z=15$. 

     We find that $>99\%$ of unresolved galaxies simulated will be too faint, given their color, to resemble a TDE as seen in Figure \ref{fig:color-color}. However, rare unresolved extremely UV-bright galaxies are possible \citep[e.g.][]{Topping2024}. In cases of excess UV emission, a flat optical continuum is still typically observed in these galaxies. A nondetection in F770W is a strong indicator that the candidate is likely not a galaxy. However, it is still valuable to obtain a spectrum or time-domain information to verify a TDE.

To summarize, our main persistent contaminants are very bright compact blue point source galaxies, local M- and T-dwarf stars, and hostless superluminous supernovae. All three contaminants can be ruled out spectroscopically. To rule out a galaxy contaminant photometrically one can use a Lyman break, and compare the absolute magnitude of the source to that of a TDE. To rule out a star contaminant photometrically, one only needs to observe the existence of a Lyman break. To rule out a superluminous supernova photometrically a second epoch of imaging is needed to observe the decline of the source.

\needspace{3\baselineskip}
\section{COSMOS-Web Search Methodology}
\label{sec:cosmos}

 COSMOS-Web is an ideal first test for this methodology. The COSMOS-Web survey images $0.54$~deg$^2$ of the sky with NIRCam to depths of $27.45$~mag in F115W, $27.66$~mag in F150W, $28.28$~mag in F277W and $28.17$~mag in F444W, and occasionally in the Mid-Infrared Instrument (MIRI) band F770W to $\sim26$~mag \citep{Casey2022}. Decades of deep multi-wavelength observations of the COSMOS field means that, for a given source, one can look back over 20 years of observations \citep[e.g.][]{Scoville2007, Capak2007, Koekmoer2011}. Furthermore, the discovery of a high redshift TDE in COSMOS-Web could be evidence of the high-redshift TDE rate enhancements that we outline in \citet{Karmen_inprep}. Given a constant TDE rate in the entire universe, \citet{Karmen_inprep} predicts $0.7$~TDEs in the COSMOS-Web dataset. However, given early-universe TDE rate enhancements discussed in Section \ref{sec:enhancements} and in \citet{Inayoshi2023}, this rate could be much higher. We apply our method to COSMOS-Web catalog \citep{Shuntov2025}, which combines new and archival imaging of the COSMOS field. Details on NIRCam image reduction can be found in \citet{Franco2025}.

\begin{figure}
    \includegraphics[width=\linewidth]{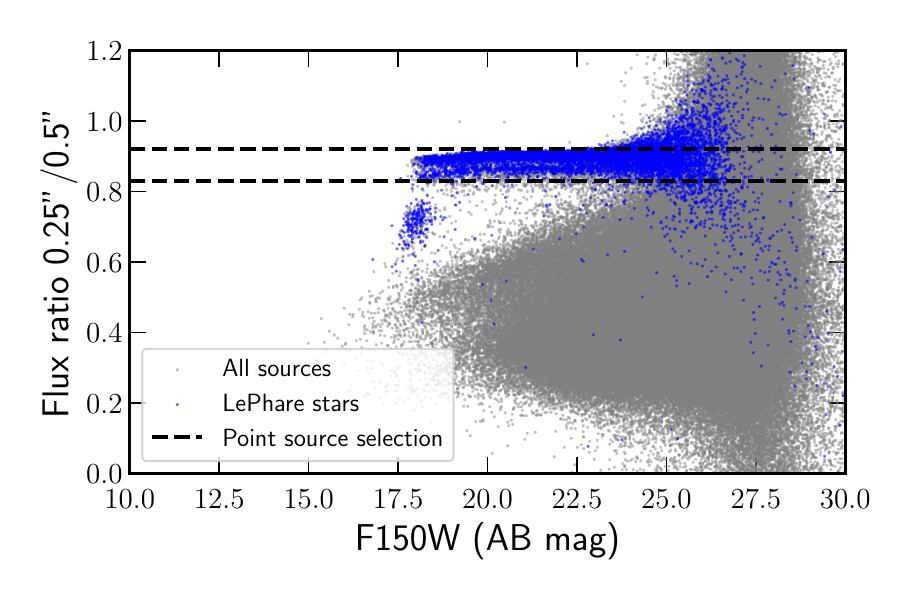}
    \caption{Flux ratio between a $0.25"$ aperture and $0.5"$ aperture in F150W vs magnitude for all sources in the COSMOS-Web catalog (gray points). Sources which are classified as stars via LePHARE SED fits (blue points) are used as a point source selection guide. The black dashed lines show the region within which we select point sources.}
    \label{fig:point_sources}
\end{figure}

To select point sources, we calculate the ratio between the flux in a $0.5$~arcsecond aperture and in a $0.25$~arcsecond aperture. We select stars by LePHARE template fitting \citep{Arnouts2011, Ilbert2006} to their SED shapes and apparent magnitudes, and find the aperture flux ratios for these sources. We calibrate our point source flux ratios to the range of flux ratios for stars, as shown in Figure \ref{fig:point_sources}.  We used the bounds of this relationship, the upper and lower limits on the flux ratio for a given magnitude, to select point sources in COSMOS-Web. We restrict only to objects which are point-like in F115W and F150W, the filters in which we most confidently expect no galaxy light beyond $z \sim 4$.

Once we have selected point sources, we use our simulated TDEs (Section \ref{sec:methods}) to define the region of color-color and color-magnitude space that high-redshift TDEs occupy. We cuts in color-color and color-magnitude relationships to restrict only to the regions where we believe high-redshift TDEs occupy. This is shown in the purple region in Figure \ref{fig:color-color}.

\subsection{Cross-match with previous COSMOS data}

    The COSMOS field has been repeatedly surveyed for the past 20 years, since the original Hubble Space Telescope COSMOS survey, which were carried out between October 2003 and November 2005 \citep{Koekemoer2007}. We use this original catalog to check the transient nature of our candidates. The HST COSMOS survey was conducted using the F814W filter, which does not overlap in wavelength with our NIRCam observations but allows an initial check for variability.

    The COSMOS field also has deep optical imaging with the Hyper Suprime Cam (HSC) on the Subaru telescope, originally in \citet{Capak2007}, and later in \citep{Tanaka2017} to $5\sigma$ AB magnitude depths of $27.8$ in the $g$-band, $27.7$ in the $r$-band, $27.6$ in the $i$-band, $26.8$ in the $z$-band, and $26.2$ in the $y$-band. These observations were conducted between March 2014 and May 2015, giving them an approximately decade-long observer-frame cadence between both the COSMOS HST ACS F814W observations, and the COSMOS-Web survey. These later HSC data comprise the COSMOS2015 catalog \citep{Laigle2016}
    
    In addition, there is previous near-infrared imaging taken in $YJHK_s$ to $\sim 26$~mag and in the narrow band $NB118$ to $24.6$~mag taken from the UltraVISTA survey, with the VIRCAM instrument on the VISTA telescope \citep{McCracken2012}. We use stacks of UltraVISTA data observed between December 2009 and June 2016. There are previous mid-infrared data from Spitzer/IRAC channel 1, 2, 3, 4 images from the Cosmic Dawn Survey \citep{Euclid2022A&A...658A.126}, which are stacks from all data taken up to January 2020. However, the IRAC data does not have sufficient angular resolution to distinguish very small, crowded sources so we cannot often use it.
    
     The COSMOS2020 catalog \citep{Weaver2022} contains all of the data mentioned (with a summary of photometry limits in \citet{Weaver2022} Table 1), and is publicly archived. In this analysis, we use forced photometry for previous surveys at the location of a detection in COSMOS-Web, and classify a nondetection as any source with a $<3\sigma$ detection in flux. Details on the forced photometry performed for all COSMOS-Web sources can be found in the COSMOS-Web catalog paper \citep{Shuntov2025}. This is useful when a TDE candidate is sufficiently bright such that a non-detection in shallower ground-based data is constraining; the detection image in COSMOS2020 is a $izYJHK_s$ stack. We use these data to extend SEDs when visualizing candidates, but do not use them in our candidate selection or fitting in case of variability between the HSC observations and COSMOS-Web.

    We search for our candidates from COSMOS-Web in the COSMOS2020 catalog using the IPAC COSMOS Archive (\href{https://doi.org/10.26131/irsa178}{doi: 10.26131/irsa178})for detections in existing COSMOS catalogs within a matching radius of 1". For the subset of candidates which are not in COSMOS2020, we download their cutouts and manually inspect them. We separate these cutouts into two groups: objects that have NIRCam F115W photometry that indicate they are above the \HST{} F814W COSMOS magnitude limit of $27.8$~mag, and those expected to be below the magnitude limit. In particular, we are interested in objects that would be bright enough to be detected in the COSMOS2020 but did not find. If a source is detected in the HST 814W mosaics, we assume that is is not a transient source. However, we do not consider a nondetection in HST confirmation of its transient nature. Although we are searching for high-redshift transients that are significantly time dilates, a $z=8$ TDE would be bright for about 3 observer-frame years, so a $\sim20$~year baseline should be enough for any high-redshift TDE to rise or fade.

\subsection{Search Results}
\label{sec:results}

    \begin{figure*}
        \includegraphics[width=\textwidth]{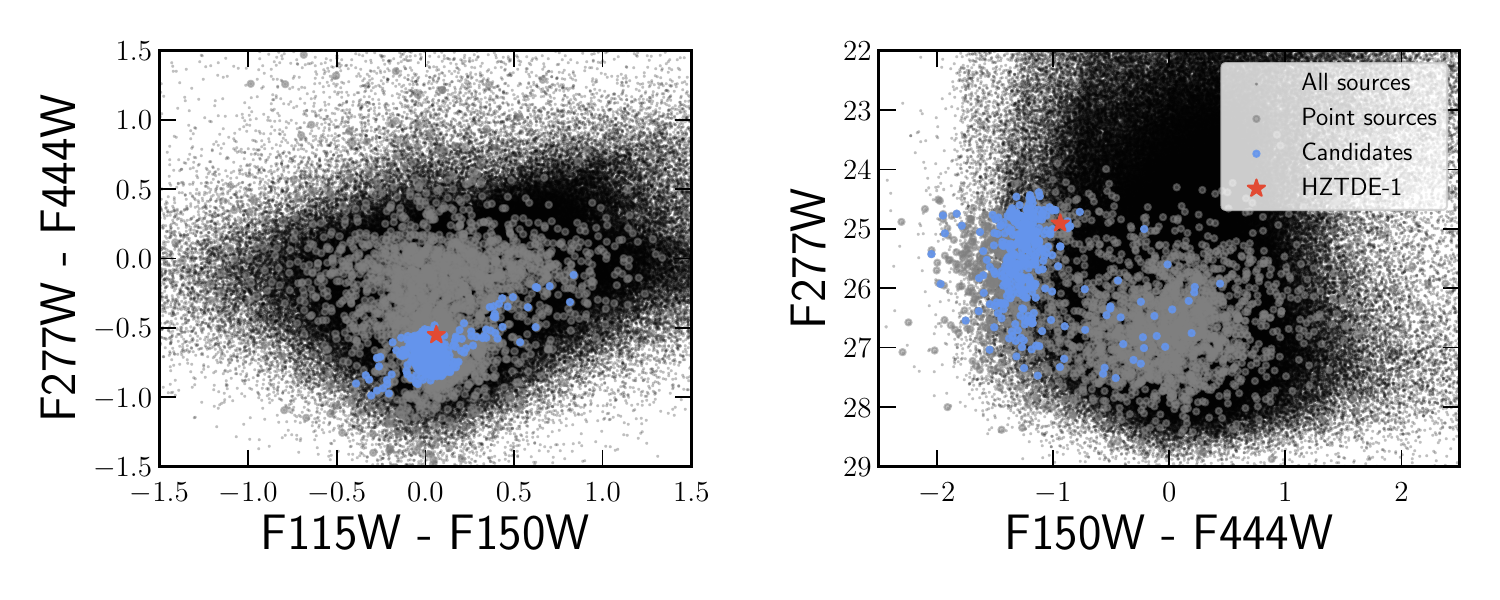}
        \caption{\textbf{Left:} Color-color diagram of all sources in black, all point sources in gray, and all selected candidates in blue points. TDE candidate HZTDE-1 is overplotted as the red star. \textbf{Right:} Color-magnitude diagram of all point sources in gray, and all selected candidates in blue points. TDE candidate HZTDE-1 is again overplotted as a red star.}
        \label{fig:selection}
    \end{figure*}

    We ultimately selected 1764 sources that passed our point source cuts, 117 passed our color-color, and magnitude cuts. We select sources based on COSMOS-Web NIRCam colors alone, in order to prevent time evolution between different surveys from affecting calculated colors. We visualize these candidates in color-color and color-magnitude space in Figure \ref{fig:selection}. 
    
    Among these sources, 82 of them had no listing in the COSMOS2020 catalog. We divide these into ``bright" and ``faint" sources. ``Bright" sources are sources with $M_{F115W} > 27.8$~AB mag, where $27.8$~mag is the detection limit in \HST{} F814W in COSMOS2020, the deepest imaging of this field. These filters do not overlap\textendash{} the $\lambda_{\rm max}$ of F814W is $9643$~\AA{} and the $\lambda_{\rm min}$ of F115W is $9975$~\AA{}\textendash{} but they are near enough to where nondetection in F115W approximates a nondetection in F814W. Therefore, ``bright" sources should have mostly been detected in \HST{}.  We find that our candidates contain 79 ``bright" sources and 3 ``faint sources". For all 79 sources without \HST{} catalog entries, we download the \HST{} F814W cutouts, and search through them by hand for nondetections. We ultimately determine by visual inspection that 72 of the 79 ``bright" sources which pass our cuts do indeed have a counterpart in F814W images, but are excluded from COSMOS2020. We note that these sources often lie on the edge of a tile, in the masked region near a bright star, or have inaccurate astrometry. Among the sources not seen in F814W in manual inspection, six of them appear to be static, yet are too faint in F814W to be detected by previous HST imaging. We are left with one source which is detected in COSMOS-Web but not in COSMOS2020. The remaining candidate peaks at $\sim25$~mag in F150W, which is $>1$~mag above the limit of previous infrared imaging with UltraVISTA. This source is not detected in UVISTA data to even $1\sigma$, confirming its nature as a transient and viable high-redshift TDE candidate. We will refer to this TDE candidate at HZTDE-1.

    \begin{figure*}
        \includegraphics[width=0.8\textwidth]{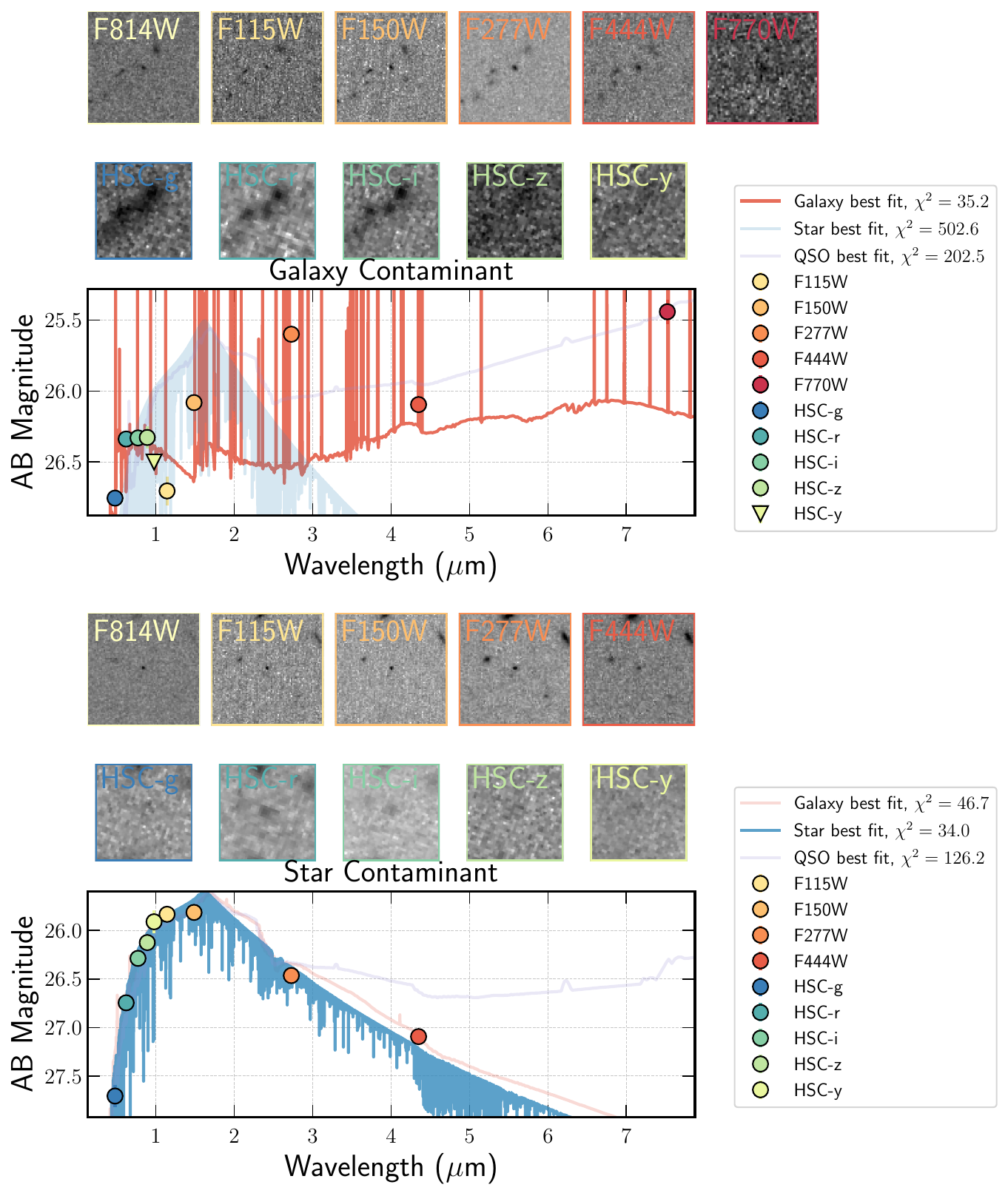}
        \caption{\textbf{Top:} Example of a high-redshift TDE candidate which we believe to be a $z=3$ galaxy contaminant. Despite having a blackbody-like SED shape in NIRCam, the HSC photometry shows that the SED is flat in the optical, with a Lyman break beginning in the g-band. Its F770W flux greatly exceeds that of a blackbody. The source appears point-like in F115W and F150W, but in the F277W imaging it can be seen to be slightly extended, although compact. The best-fit galaxy template is plotted in red, star template in light blue, and quasar in purple. This galaxy does not have a F814W flux measurement in the COSMOS2020 catalog due to photometry flags, but manual inspection confirms it was detected. \textbf{Bottom:} Example of a high-redshift TDE candidate which we believe to be a local cool dwarf star. The source has no listing in the Hubble Source Catalog, although it is detected in COSMOS2020. This source has no Lyman break, and is detected to $5\sigma$ in every HSC band.}
        \label{fig:contaminant-plots}
    \end{figure*}

    These candidates are selected only by \JWST{} photometry, and the HSC photometry is consistent with a blackbody with the same peak as a high redshift TDE for the vast majority of sources, as seen in Figures \ref{fig:contaminant-plots} and \ref{fig:cand-tde}. Figure \ref{fig:cand-tde} shows an example fit of our simplified high-redshift TDE model to HZTDE-1, characterized only by absolute magnitude, blackbody temperature, and redshift, to the photometry. We find that, while the model fits well to the majority of our data, the degeneracy between redshift and temperature means that the fits are extremely dependent on the starting guess and there is no way to discern a source's redshift without observing a Lyman dropout. Additionally, shown as a comparison, this could also match the SED of an M-dwarf star. Therefore, for these data it is essential that we utilize all previous \HST{} imaging to check for variability and optical HSC data to look for a dropout.

    Cross-identification in the Hubble Source Catalog suggests that candidates that have Lyman dropouts are likely high-redshift galaxies. In the \HST{} COSMOS catalog, many of our faint candidates are labeled ``high-redshift galaxy", some with spectra confirming these labels. This is explained by our point source selection: we select point sources based on their flux ratios in two small apertures (See Section \ref{sec:cosmos}), rather than doing actual morphological fitting. This means that a sample of slightly lower-redshift, brighter and bluer galaxies could contaminate the sample by having an approximate angular size of order of the smaller aperture we use, rather than a smaller angular size than the NIRCam PSF. Better morphology selection could reduce this contamination, but nothing can eliminate rare perfect contaminant compact, blue, high-redshift galaxies other than time-evolution. We show an example of these high-redshift galaxy contaminants in Figure \ref{fig:contaminant-plots}. HSC photometry confirms that it is indeed not a blackbody, and has a Lyman dropout in the g-band. Additionally, while it appears point-like in F115W and F150W, the F277W imaging shows that it is an extended source. The object is just under a $5\sigma$ detection in the HST image, but s visible in the cutout. We also search all of our objects in the Chandra X-ray imaging of the COSMOS field, and find no X-ray sources associated with any object \citep{Civano2016}. This is expected for a high-redshift TDE, because their soft X-ray spectrum would be completely absorbed by the CGM beyond $z\sim 3$.

    Objects without Lyman dropout are likely cool dwarf stars. Stars with temperatures $2400 \textrm{K} < T < 3500 \textrm{K}$ have blackbody peaks around 1.2~$\mu$m, the same observed wavelength as a $z > 5$ TDE. These stars are abundant in the Milky Way, although are less common when looking out of the Galactic Plane. Figure \ref{fig:contaminant-plots} shows a source which we believe is a star. It has no listing in the COSMOS2020 catalog, but manual inspection can identify the corresponding object in previous COSMOS imaging.

\section{Properties of the TDE Candidate HZTDE-1}
\label{sec:candidate}

    \begin{figure*}
        \centering
        \rotatebox{90}{
        \includegraphics[width=0.9\textheight]{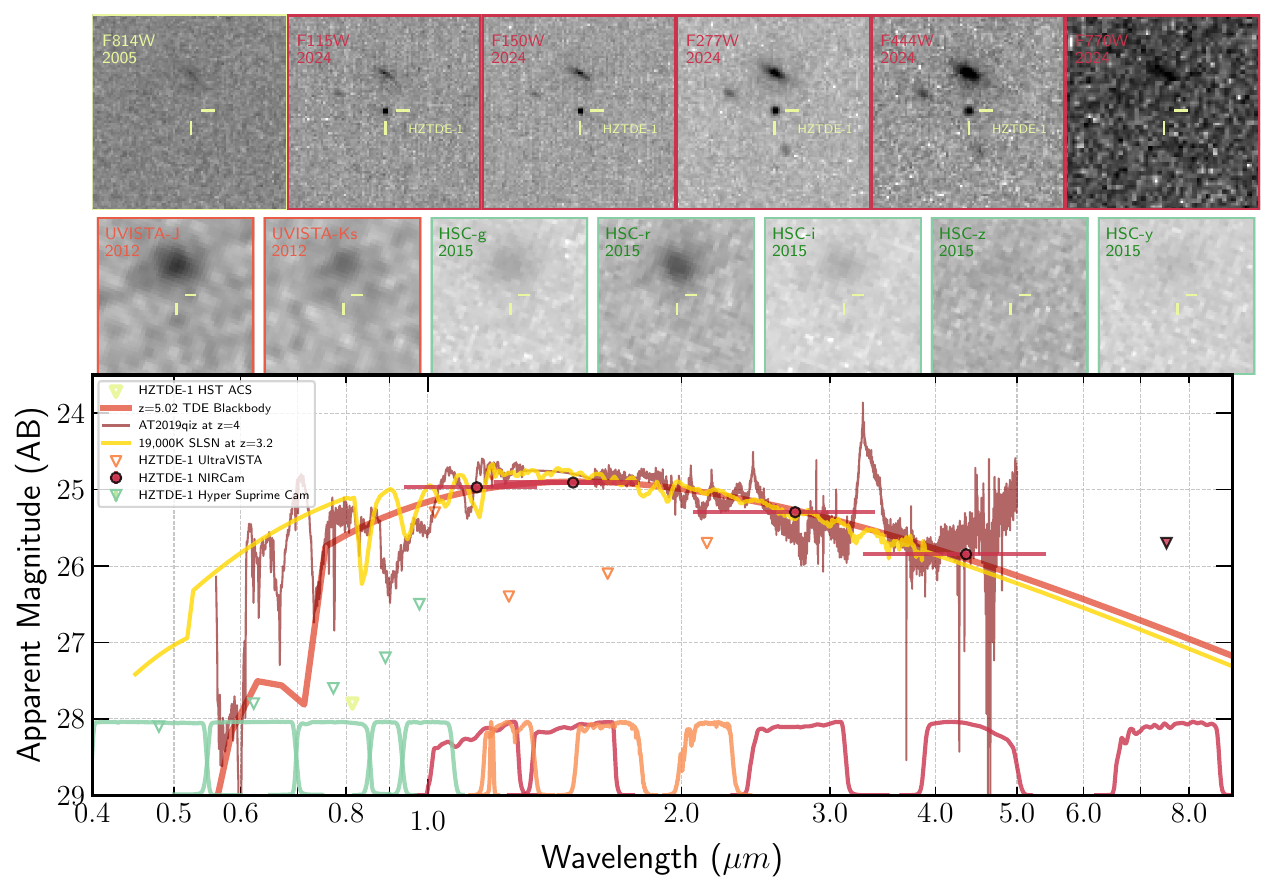}
        }
        \caption{Cutouts (top panels) and SED (bottom panel) of high redshift TDE candidate, HZTDE-1. The cutouts include imaging from UltraVISTA, Hyper Suprime Cam, Hubble ACS, and \JWST{} NIRCam. The source is only detected to $>3\sigma$ in the NIRCam images. The SED shows the observed SED as red circles, the best fit SLSN spectrum (blue line) and TDE spectrum (red line), and all photometric upper limits as downward-pointing triangles. Errorbars in magnitude exist for the NIRCam data but are too small to be seen. The transmission curve for each filter is plotted below it. A simulated spectrum of TDE 2019qiz at $z=4$ is overplotted as a dotted line.}
        \label{fig:cand-tde}
    \end{figure*}

The TDE candidate HZTDE-1 is shown in Figure \ref{fig:cand-tde}. This source was the only object which is a point source in NIRCam, has no optical counterpart, has the SED shape of a high-redshift blackbody, and does not appear in the COSMOS2020 catalog. Furthermore, HZTDE-1 is not detected in previous deep UltraVISTA imaging, which verifies the transient nature of the object. Due to possible contamination from a nearby galaxy, we verify the nondetection in HSC and UltraVISTA by performing aperture photometry to manually compare the flux at the location of HZTDE-1 to the background flux. We describe this procedure in Appendix \ref{sec:force_phot}, and find that there truly is no detection above background in any previous optical or NIR imaging. LePHARE spectral template fitting does not settle on a good classification for this source given its SED: its reduced chi-squared statistic (per degree of freedom) for the best star template is $\chi_\nu^2 = 477$, for an AGN template is $\chi_\nu^2 =1050$, and $\chi_\nu^2 =401$ for the best galaxy template, which fits a galaxy at $z=7.33$. We also show in Figure \ref{fig:cand-tde} a composite spectrum of TDE 2019qiz, which is constructed by connecting an ultraviolet spectrum taken with HST UVIS \citep{Hung2021} and an optical spectrum taken with the X-Shooter instrument on the Very Large Telescope \citep{Nicholl2020}. We select this TDE due to its broad wavelength coverage near maximum brightness; it best fits the HZTDE-1 photometry at $z=4.1$ due to its lower blackbody temperature ($15,000$~K), but we simulate its spectrum at $z=5$ to show the locations of common TDE emission lines.

\begin{figure*}
        \centering
        \includegraphics[width=\textwidth]{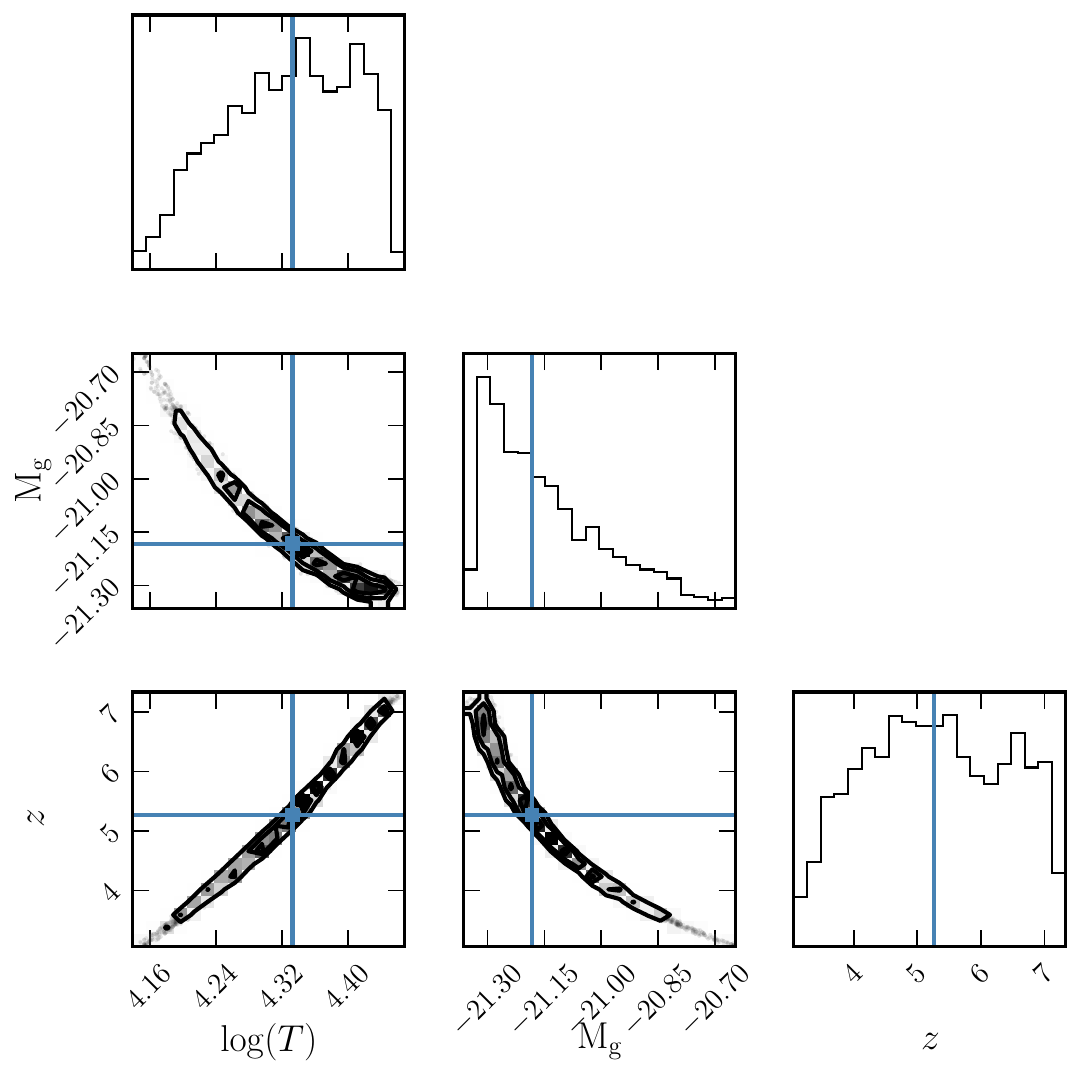}
        \caption{MCMC samples of the posterior distribution for each parameter in the TDE model fit to the HZTDE-1 photometry. Errorbars are wide due to degeneracies between temperature, redshift, and luminosity. We use the observed temperatures and maximum luminosities of ZTF TDEs as priors. We also set bounds on the redshift fit as decribed in Section \ref{sec:candidate}.}
        \label{fig:mcmc_fit}
    \end{figure*}

We approximate its redshift through maximum a posteriori fitting of our blackbody TDE model with Gaussian uncertainties optimizing through Monte Carlo Markov Chain sampling. We utilize the ZTF TDE sample's distribution of TDE temperatures and luminosities as priors. Our prior on redshift is a uniform distribution between $z=3.5$ and $z=7.5$. We estimate a lower bound on the redshift through our simulations of TDE hosts: we expect that the lowest-mass TDE host would not be detected in the two short-wavelength NIRCam filters of COSMOS-Web at $z\geq3.5$. We can set a strict upper limit on the redshift because at $z=7.5$, the F115W emission would have been Lyman absorbed. We find our most likely TDE to have $z=5.02^{+1.33}_{-1.11}$, $M_g = -21.15^{+0.21}_{-0.13}$, and $\log(T_{BB})=4.31^{+0.09}_{-0.09}$. We show the MCMC sampling of the posterior distribution in Figure \ref{fig:mcmc_fit}. If HZTDE-1 were as cool as the lowest-temperature known TDEs, $\sim 10^4$~K, it could be as low-redshift as $z=2$; however, as $z=2$ a typical $10^{10}~M_{\odot}$ host galaxy would be resolved in COSMOS imaging.

\subsection{Comparison to supernova models}
\begin{figure*}
    \centering
    \includegraphics[width=\linewidth]{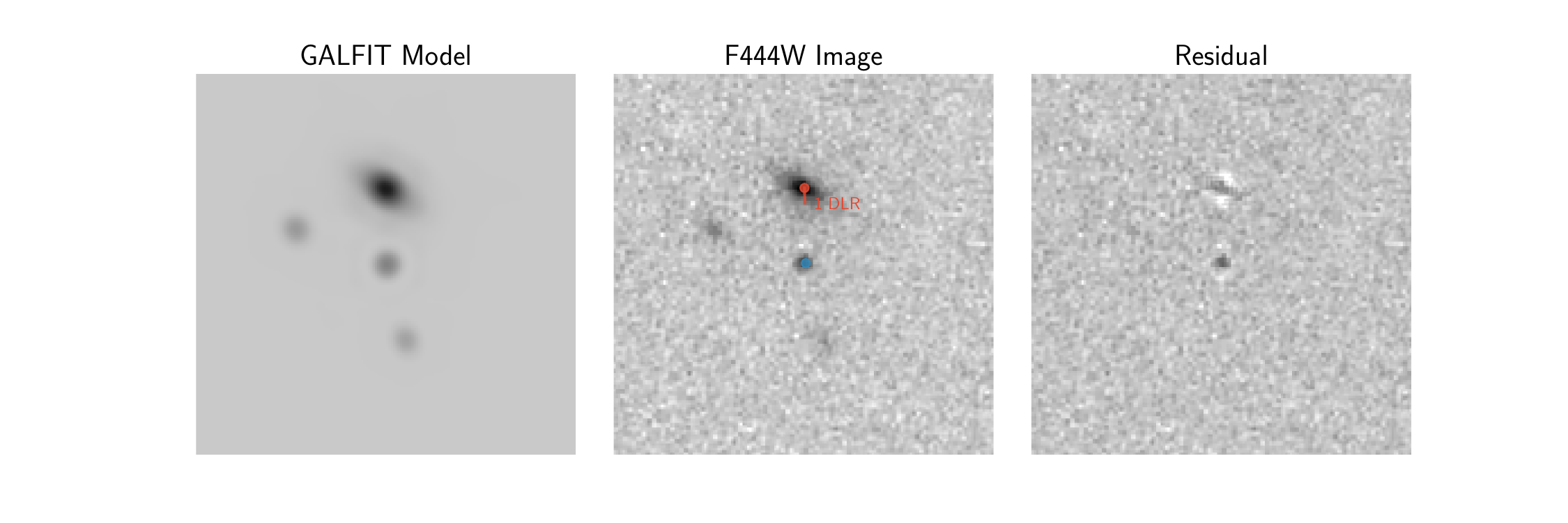}
    \caption{GALFIT modeling of the morphology of the sources in the F444W imaging of HZTDE-1. We display the size of $1$ directional light radius (DLR) from the galaxy to the transient to show the accuracy of the modeling. The HZTDE-1 is $>5$ DLR from the galaxy, unlikely to be associated. In shorter-wavelength filters, this DLR separation increases.}
    \label{fig:galfit}
\end{figure*}

There are three nearby galaxies that HZTDE-1 could belong to if it is a lower-redshift supernova. We assess its possible association with the galaxy directly above it in the cutouts shown in Figure \ref{fig:cand-tde}, because it is the closest galaxy, with an angular separation of $0.998$", the brightest galaxy with $M_{F444W}= 23.7$~AB~mag, and the most extended with effective radius $R_e = 0.20$" as modeled by GALFIT \citep{Peng2002}. The GALFIT model for the sources in the field is shown in Figure \ref{fig:galfit}, alongside the F444W image and the residual. The best-fit galaxy spectrum to this source's SED has a photometric redshift $z_{\rm phot} = 1.75 ^{+0.1}_{-0.07}$.  This galaxy has a slightly above average best-fit specific star formation rate given its photometric redshift, of $\log(\rm{sSFR}) / M_\odot~yr^{-1} = -8.67$. An especially high star formation rate would support the occurrence of a supernova in its outskirts \citep{Botticella2012}. The SED is well-sampled from the HSC $g$-band to NIRCam F444W, and the model spectrum fits accurately with a $\chi^2=1.72$ per degree of freedom. We follow the technique developed in \citet{Gupta2016} to assess host galaxy association for a given supernova through calculating the directional light radius (DLR), which is the elliptical radius of a galaxy in the direction of the supernova. They then calculate the $d_{\rm DLR}$, which is the distance between the supernova and the galaxy normalized by the DLR. They utilize both $HST$ ACS data and MICECATv2.0 simulations to show that any SN with a galaxy $>5$~DLR away should be assigned ``hostless", and achieve $\sim 95\%$ completeness with a similar cut. We calculate the unitless dDLR for HZTDE-1 to be $5.05$ relative to the nearest galaxy, making it highly likely to be a hostless transient. This dDLR is even higher for the short-wavelength NIRCam filters ($>6$~DLR in F115W), which have higher spatial resolution.

\begin{figure*}
    \centering
    \includegraphics[width=\linewidth]{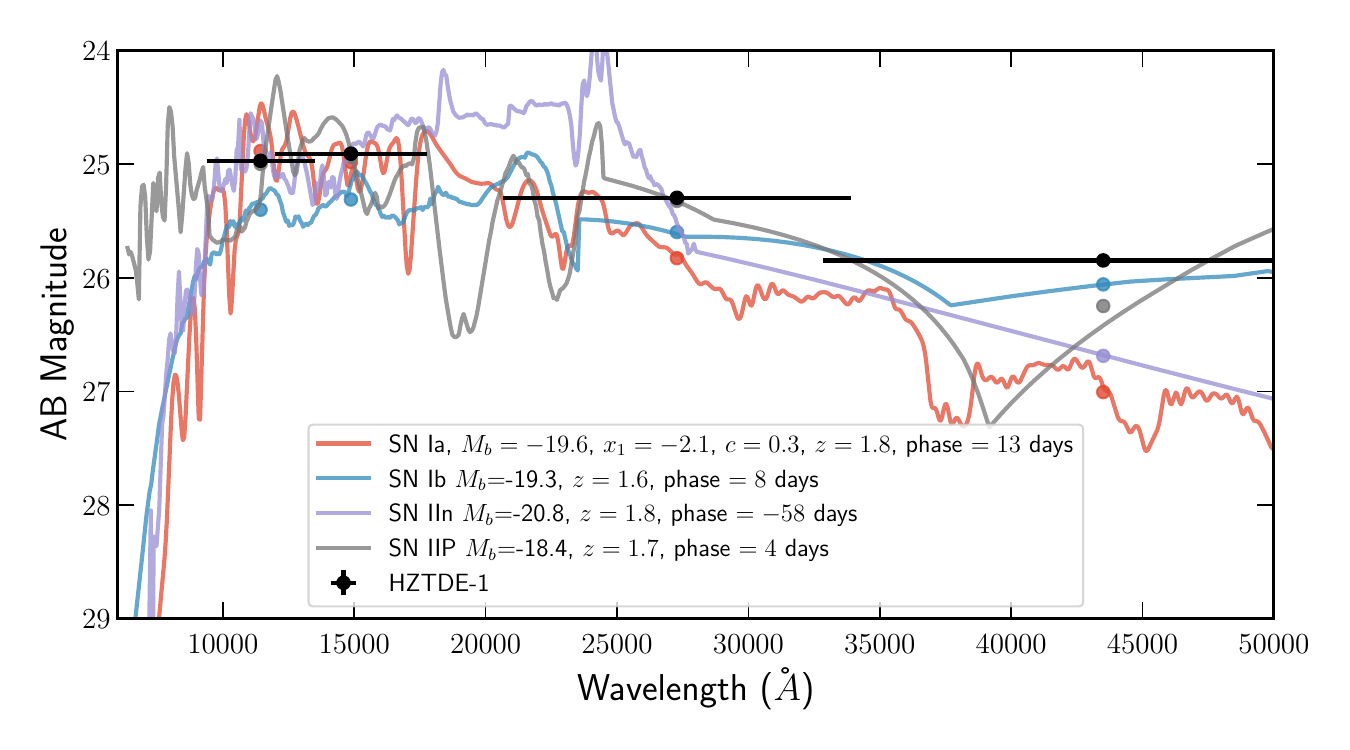}
    \caption{Modeled spectra and synthetic photometry of best-fit Type Ia, Type Ib/c, IIP, and IIn supernovae compared to the SED of HZTDE-1. All SNe are within $z=1.75\pm0.1$, in order to be associated with the most likely host galaxy. Type Ia and Ib supernova cannot successfully fit the SED. A very bright IIP or IIn could fit the majority of the SED, but struggle to produce the same infrared luminosity.}
    \label{fig:supernovae}
\end{figure*}

\begin{figure*}
    \centering
    \includegraphics[width=\linewidth]{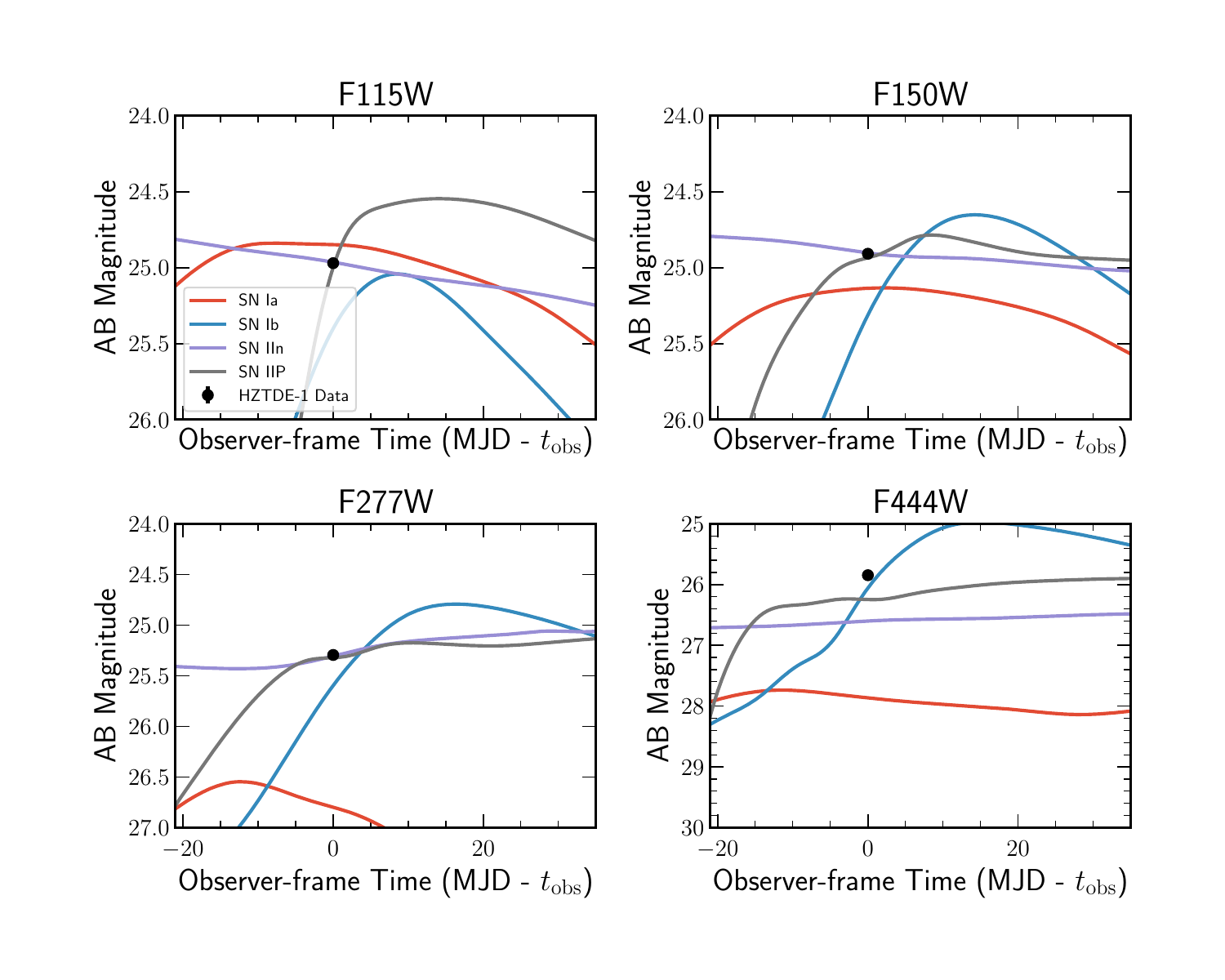}
    \caption{Lightcurves of best-fit supernovae models to HZTDE-1, given that they are at redshift within $0.1$ of the closest galaxy. Each panel is in a single COSMOS-Web NIRCam filter. The best-fitting models, IIP and IIn, do not adequately explain the F444W flux, which would be at $\sim1.6\mu$m in the rest frame.}
    \label{fig:lightcurves}
\end{figure*}

We assume that HZTDE-1 is unlikely to be a ``hostless" non-superluminous supernova. However, given the rare case that HZTDE-1 is associated with the nearest galaxy despite extreme distance, we fit supernova SEDs for each supernova subtype at its photometric redshift, $z=1.75\pm0.10$. We use the full range of supernova properties described in Section \ref{sec:contaminants}. For every supernova subtype, we fit every available model in \texttt{sncosmo} \citep{sncosmo}, extending to the infrared with a blackbody when the model has insufficient wavelength range. We do not fit models for Population III SNe, Pair-Instability Supernovae (PISNe), or other exotic transients. We show our best-fit supernova spectra in Figure \ref{fig:supernovae}, and their respective lightcurves in Figure \ref{fig:lightcurves}. We find that HZTDE-1 is significantly too red to be a SN Ia at $z\sim1.75$ and too bright to be a SN Ia at $z>2$. HZTDE-1 is $\gtrapprox1$~mag brighter than a SN Ib/c at $z\sim1.75$. The SED is somewhat plausibly explained by an anomalously bright IIP before maximum brightness or a bright IIn prior to maximum brightness, but both under-predict the flux in F444W at $z\sim1.75$, which would be $1.6~\mu$m in the rest-frame. These best models are shown in Figure \ref{fig:supernovae}. This excess flux could be explained by contribution from thermal dust emission, but given the extreme offset from the potential host it is unlikely that the supernova would be embedded in dust. In summary, for HZTDE-1 to be a supernova associated with its nearest galaxy it must be an anomalously bright SN IIn at an extreme offset with an infrared flux excess.

There is a possibility that HZTDE-1 is a superluminous supernova (SLSN), as shown in Figure \ref{fig:cand-tde}. SLSNe have similar intrinsic luminosities to TDEs and their spectral continua are primarily blackbody thermal emission \citep{Gomez2024}. Their host galaxies tend to be less massive, meaning that at large enough distances ($z\gtrsim 3$) they may also outshine their hosts and be point sources. Therefore, to be certain whether HZTDE-1 is a high-redshift TDE or high-redshift hostless supernova, either follow-up spectroscopy or monitoring of its photometric evolution is necessary. With a single photometric detection, the relative rates of TDEs and SLSNe at $z>3$ is the main factor that determines what HZTDE-1 likely is. Current best estimates of the volumetric rate of SLSNe place them somewhere between $0.91 \times 10^{-7}~\rm{SNe}/\rm{year}/\rm{Mpc}^3$ at $z=1$ \citep{Prajs2017} and $1.99 \times 10^{-7}~\rm{SNe}/\rm{year}/\rm{Mpc}^3$ at $z=0.2$ \citep{Quimby2013}, as compared to the current best estimate for the local optical TDE volumetric rate of $3.1 \times 10^{-7}~\rm{TDEs}/\rm{year}/\rm{Mpc}^3$. \citet{Tanaka2013} predicts the volumetric rate at SLSNe at higher redshifts (up to $z\geq 10$) by scaling it with the cosmic star formation rate (SFR) density using two SFR models \citep{Robertson2012, Hopkins2006}. They calibrate to the SLSN low-redshift rate, and account for the initial mass function and the fraction of massive stars to find that the SLSN volumetric rate increases with redshift until $z\sim 5$, where the SFR density peaks, and then the rate decreases with redshift indefinitely. At its maximum, $z\sim5$, the volumetric SLSN rate is $10\times$ the local rate. We expect the TDE rate to also be elevated by at least $10\times$ the local rate, as shown in \citet{Karmen_inprep}. HZTDE-1 would need to be blue for a
SLSN, with a best-fit temperature 24, 000 K at best-fit
redshift z = 3.6, but this is possible briefly before peak
and during peak brightness. However, this fit was performed without a prior on SLSN behavior and a more reasonable 19,000 K SN will fit at z=3.2. Indeed, because of the low-mass nature of SLSN hosts, the SN could occur in a lower redshift, very faint host galaxy and would be intrinsically redder.  Given the sample of host galaxies for the Palomar Transient Factory observed SLSNe \citep{Perley2016}, these hosts could drop out of the COSMOS-Web survey as soon as $z\sim1.7$. This is around the peak of cosmic star formation rates and would imply a reasonable number of hostless SLSNe. Therefore, without further monitoring, the possibility of a lower-z SLSN is equally likely to a TDE.

\section{Discussion and Future Confirmation}
\label{sec:discussion}

In this work, we model the properties of TDEs at high redshift assuming that their populations are similar to those at low redshift. We argue that the assumptions made about their host galaxies (Section \ref{sec:hosts}) and TDE spectra (Section \ref{sec:high_z_tde}) set limits on the colors and morphologies of high-redshift TDEs. However, any number of these assumptions can be relaxed. In future work, for example, modeling of TDEs in extended hosts, or modeling of dust-extincted TDE emission can increase the efficiency of searches for TDEs in wide-field surveys. As this search offers an opportunity to study the UV emission shortward of $2000$~\AA{} which is usually not observed, deviations from blackbody assumptions about TDE spectra, such as those mentioned in Section \ref{sec:other_model} may emerge from studies of high-redshift TDEs.

Either spectroscopic or photometric follow-up could be used to confirm HZTDE-1's classification as a TDE and its redshift. TDEs have blue spectra with blackbody continua and broad Hydrogen and/or Helium emission features that narrow as the TDE evolves. A key distinction between TDEs and supernovae is their decline: TDEs do not change temperature as they cool \citep{vanVelzen2020}, while supernovae cool rapidly after peak, typically 5000-8000~K \citep{Gomez2024, Aamer2025}. Without imaging of a host galaxy to verify that HZTDE-1 is nuclear, either identification of spectroscopic features or photometric monitoring is needed to confidently confirm it as a TDE. However, because of the degeneracy between redshift, luminosity, and temperature, a spectrum of the TDE or host would be needed to calculate a redshift. Spectroscopy, specifically in the infrared with \JWST{}'s NIRSpec, is key to observing common TDE emission lines. X-ray emission is not expected because, as mentioned, the CGM will absorb all of the soft X-ray emission characteristic of TDEs at $z>3$. Longer wavelengths may be worth monitoring for later-time infrared flares due to dust echoes \citep{vanVelzen2016}.

If HZTDE-1 is a high-redshift SLSN, its discovery and classification will be valuable for understanding stellar evolution at higher redshift. SLSNe are expected to result from the explosions of massive stars \citep[up to tens of solar masses][]{Blanchard2020} and have a strong preference for galaxies with low metallicities \citep{Perley2016, Schulze2018, Cleland2023}. To first order, the cosmic rates of SLSNe should follow the cosmic star formation rate \citep{Tanaka2013}. Deviations from this relationship could provide insight into the abundance of metal-poor galaxies at earlier cosmic times. The methodology outlined in this paper could identify potentially hundreds of high-redshift SLSNe in upcoming wide-field surveys (e.g. the Roman High Latitude Wide Area Survey) and therefore could lead to measurements of how their rates and hosts change with redshift. This in turn can be used to study how massive stars form and evolve as the metallicity of the universe changes over cosmic time.

\subsection{TDE rates in COSMOS-Web}
\label{sec:enhancements}
In the companion work \citet{Karmen_inprep}, we calculate observed TDE rates as a function of redshift in LSST, Roman, and the COSMOS-Web survey. We find that if we simply extrapolate the local TDE rate, scaled by the number density of SMBHs \citep{Shankar2009} that can disrupt a main-sequence star \citep[following][]{Kochanek2016, Sun2015, Donnarumma2015}, we get a $20\%$ chance of finding a single TDE in the COSMOS-Web survey. 

We show, however, that the observed properties of high-redshift galaxies would imply a dramatically enhanced rate of TDEs at high-redshift $z>5$, and perhaps as low as $z=2$. Namely, many galaxies are observed to be centrally compact \citep{Finkelstein2023, Kartaltepe2023, Baggen2023, Guia2024}, and may host nuclear stellar clusters \citep{Ricotti2016, Sun2023} at high redshifts. This increases central stellar density, $\rho_{\star}$, by $1-2$ orders of magnitude \citep{Baggen2024a} which in turn can easily lead to a $\times 100$ TDE rate enhancement assuming a galaxy's TDE rate scales as $\propto \rho_\star^{0.65}$ \citep{Pfister2020}, or even higher given the more dramatic observed scaling with stellar density $\propto \rho_\star^{0.9}$ \citep{Graur2018}. This enhancement is in agreement with simulations of TDEs in galaxies with nuclear stellar clusters \citep{Kritos2024, Pfister2020}. Additionally, galaxy mergers are shown to enhance TDE rates both observationally and theoretically. Host studies of optical/UV TDEs demonstrate a clear preference for galaxies which are believed to be products of mergers \citep{French2020, Wevers2024}, and the TDE rate is seen in hydrodynamical simulations to be enhanced by an order of magnitude for 100~Myr after a merger \citep{Pfister2021}. With a galaxy merger rate that increases as a function of redshift, peaking at cosmic noon \citep{Ventou2017, Duncan2019, Perna2025} we predict TDE rates to be enhanced from mergers as well. In previous works \citep[e.g.][]{Kochanek2016, Donnarumma2015}, the volumetric TDE rate has been scaled with a quickly-evolving SMBH mass function, as as a result it was expected that TDE rates decrease rapidly with the SMBH mass function as a function of redshift. However, given the abundance of supermassive and overmassive black holes detected at early cosmic times \citep[e.g. ][]{Maiolino2023a, Larson2023bh, Kokorev2023}, this phenomenon will be significantly reduced. TDEs are among the few luminous phenomena very close to an SMBH, and are believed to produce super-Eddington accretion at early times \citep{Strubbe2009} so their study at high redshifts can give an indication to the mechanisms by which early SMBHs accrete and grow.

The TDE rate can still be enhanced at high redshifts without the detection of an overabundance of TDEs in \JWST{} imaging. Samples of infrared-bright TDEs find that the majority of TDEs with no detected optical/UV emission occur in star-forming galaxies \citep{Masterson2024}. It is known that the fraction of star-forming galaxies increases with redshift beyond $z>2$ \citep{Madau1996}. Therefore, if the majority of high-redshift TDEs occur in star-forming galaxies, it follows that their UV emission may be dust obscured, reprocessed, and emitted as rest-frame infrared emission. This possibility is supported in \citet{Inayoshi2023} which considers the enhancement of the rate of TDEs in obscured AGN. In order to detect the remainder of high redshift TDEs, if they reside in star-forming galaxies, one may need to look for the rest-frame infrared dust echoes. This rest-frame ultraviolet search is a first probe of these rate enhancements. In order to confirm the classification of HZTDE-1, spectroscopic follow-up must be done before the source completely fades.

\section{Conclusion}
\label{sec:conclusion}
In this work we search the COSMOS field for ongoing high-redshift TDEs, and successfully identify one $z\sim5$ TDE candidate. We summarize our findings below:
\begin{itemize}
    \item We show that beyond $z>4$, given both the nature of TDE preferred host galaxies and the low masses of galaxies at high redshift, the high-redshift counterparts to local, optical/UV TDEs will be hostless point sources. We demonstrate that they will be k-corrected into the wavelength range of \JWST{}'s NIRCam and could be detected to $z\sim10$.
    \item We develop criteria for identifying candidate $z>4$ TDEs given their NIRCam colors, comparing to contaminant stars, galaxies, AGN, and supernovae. For regions of potential contamination, we pose methods for distinguishing TDEs from these contaminants. This methodology can be applied to any deep field survey, and will yield a wealth of high-redshift TDEs from the Roman High Latitude Wide Area Survey. This methodology could also be used to identify a range of other hostless transients in upcoming surveys.
    \item We apply this selection method to the COSMOS-Web catalog, and identify 117 sources among $>700,000$ that pass our cuts. We systematically check these sources for lack of variability using the COSMOS2020 catalog, which compiles all previous imaging of the COSMOS field, and find one source which is a confirmed transient. We call this TDE candidate HZTDE-1.
    \item We model HZTDE-1 as a possible supernova, both through SED modeling and host galaxy association. We find that HZTDE-1 is $>5$ directional light radii from the nearest galaxy, making an association extremely unlikely. We find that while the SED cannot be explained by a Type I supernova, a Type II (particularly IIn) at $z\sim2$ can explain the SED mildly well, except for the farthest infrared emission. However, it is still possible that this transient is a ``hostless" luminous or superluminous Type IIn supernova.
    \item We model HZTDE-1 as a possible TDE, and find that the most likely TDE has an effective blackbody temperature of $\log(T_{\rm{BB}}) = 4.31\pm 0.09$, absolute g-band magnitude of $M_g= -21.15^{+0.21}_{-0.13}$ and redshift of $z=5.02^{+1.33}_{-1.11}$.
    
\end{itemize}

Follow-up observations of HZTDE-1, for photometric fading and for spectroscopic information, are crucial for understanding its nature and confirming its redshift. If HZTDE-1 is a TDE, it could confirm an enhancement of the volumetric rate of TDEs in the high-redshift universe. As TDEs offer a probe of central SMBH mass independently from the virial methods that use AGN emission, a population of TDEs at high redshifts could offer insight into the nature of the SMBH mass function as a function of cosmic time. This, in turn, could place additional constraints on the methods by which SMBHs seed and grow.

\section{Acknowledgements}

The authors thank Christina Lindberg for discussions about dust, and Sean Carroll for his discussion of source selection. We also thank Muryel Guolo for his TDE wisdom and cynicism. This paper is based in part on observations with the NASA/ESA Hubble Space Telescope and James Webb Space Telescope obtained from the Mikulski Archive for Space Telescopes at STScI. This research has made use of the NASA/IPAC Infrared Science Archive, which is funded by the National Aeronautics and Space Administration and operated by the California Institute of Technology. This material is based upon work supported by the National Science Foundation Graduate Research Fellowship under Grant No. DGE2139757. The French contingent of the COSMOS team is partly supported by the Centre National d’Etudes Spatiales (CNES). OI acknowledges the funding of the French Agence Nationale de la Recherche for the project iMAGE (grant ANR-22-CE31-0007). This work was made possible by utilising the CANDIDE cluster at the Institut d’Astrophysique de Paris. The cluster was funded through grants from the PNCG, CNES, DIM-ACAV, the Euclid Consortium, and the Danish National Research Foundation Cosmic Dawn Center (DNRF140). It is maintained by Stephane Rouberol.

\clearpage
\appendix
\section{Verification of optical nondetection}
\label{sec:force_phot}

\begin{figure*}[!h]
    \centering
    \includegraphics[width=\linewidth]{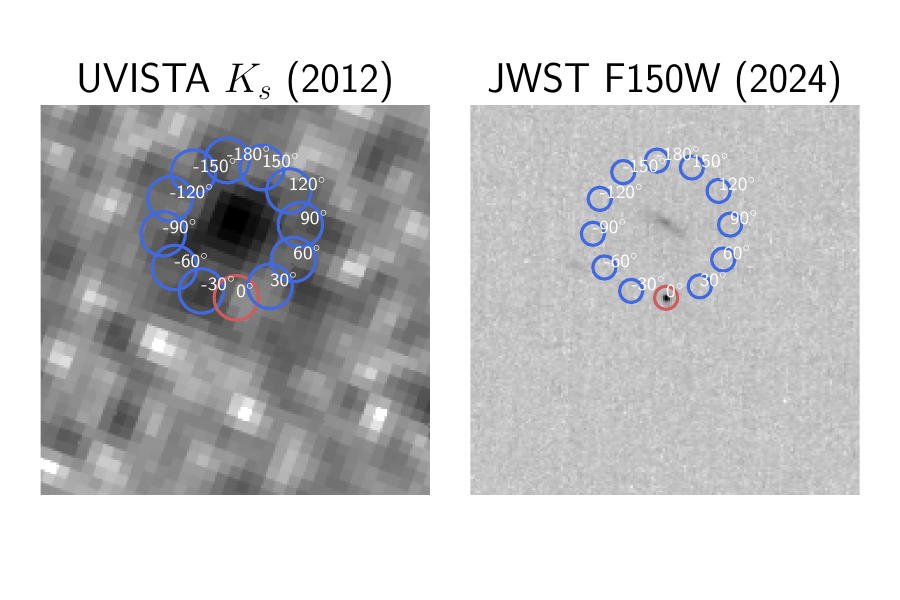}
    \caption{\textbf{Left:} UltraVISTA $K_s$-band imaging of galaxy near HZTDE-1, surrounded by the apertures described in Appendix \ref{sec:force_phot}. The red aperture is centered on HZTDE-1. \textbf{Right}: \JWST{} NIRCam F150W imaging of the same galaxy and HZTDE-1, surrounded by apertures in the same location. Again, visibly, the red aperture contains HZTDE-1. For visualization purposes and because \JWST{} has a very high angular resolution, the \JWST{} apertures are slightly smaller.}
    \label{fig:aperture_plot}
\end{figure*}

\begin{figure}
    \centering
    \includegraphics[width=\linewidth]{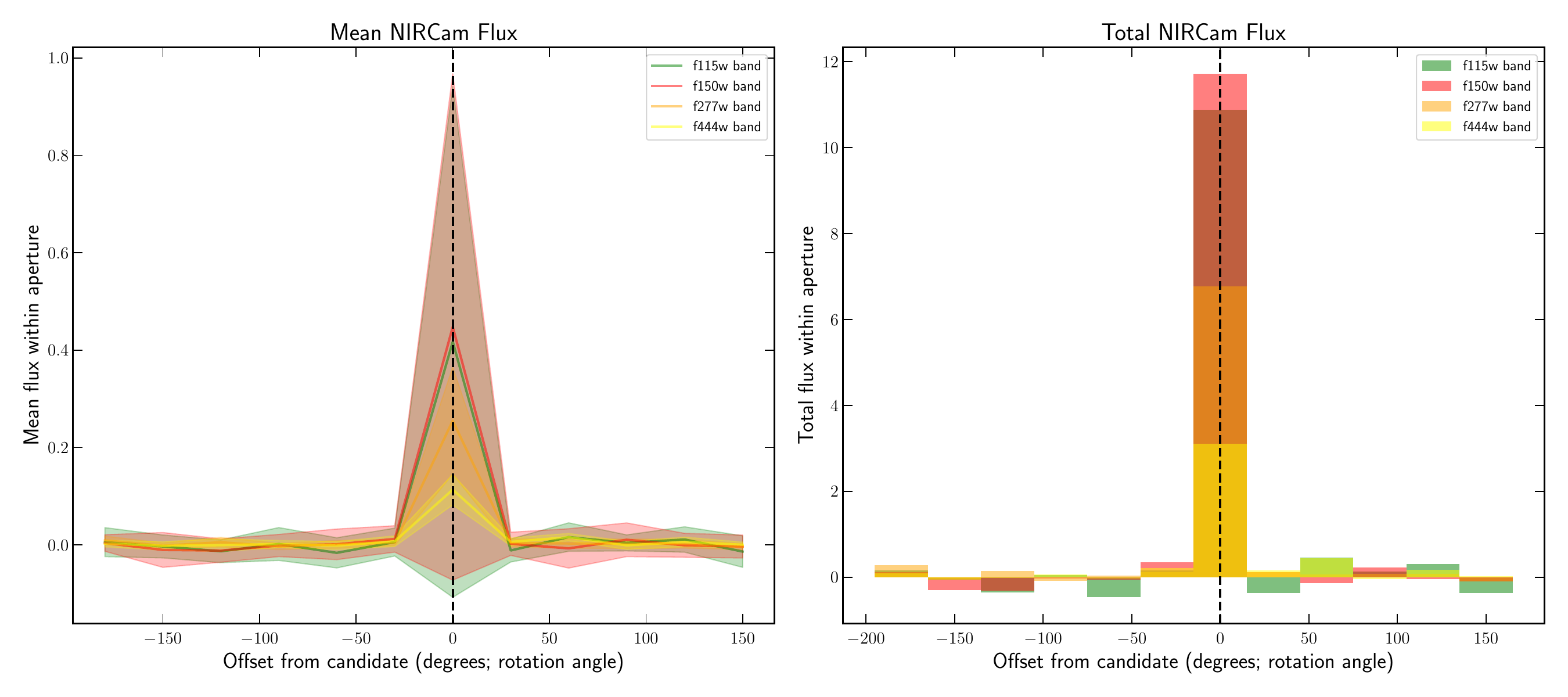}
    \caption{Aperture from \JWST{} imaging using apertures in the same locations as for UltraVISTA relative to the nearby galaxy. For visualization purposes and because \JWST{} has a very high angular resolution, the apertures are slightly smaller. $0^\circ$ corresponds to the location of HZTDE-1. Using NIRCam, there is a clear detection of HZTDE-1 with respect to the background.}
    \label{fig:jwst_aperture}
\end{figure}

Because of the small angular distance between HZTDE-1 and the nearby galaxy, we wanted to verify that HZTDE-1 was not obscured by the galaxy in the lower-resolution UltraVISTA imaging. We use aperture photometry to test this as follows. First, in pixel space, we calculate the location of the centroid of the galaxy in each filter by fitting a 2D gaussian to the data in the \texttt{photutils} Python package \citep{larry_bradley_2025_14889440}. Filter by filter, we find the angular separation between the galaxy centroid and the PSF-modeled coordinates of HZTDE-1 which we call $d$. Starting with the location of HZTDE-1, we place a series of apertures, each with a radius of $2$~pixels, at an angular distance $d$ from the galaxy, each offset from each other by a rotation angle of $20^\circ$, as shown in Figure \ref{fig:aperture_plot}. We calculate both the mean flux and the total flux within each aperture in UVISTA and find no increase in the flux at the location of HZTDE-1 compared to background in any filter. We then perform the same procedure using \JWST{} imaging, with apertures of radius 3 pixels, as seen in Figure \ref{fig:aperture_plot}. As expected, we find a $\times 1000$ increase in flux relative to background in NIRCam, as seen in Figure \ref{fig:jwst_aperture}.

\begin{figure}
    \centering
    \includegraphics[width=\linewidth]{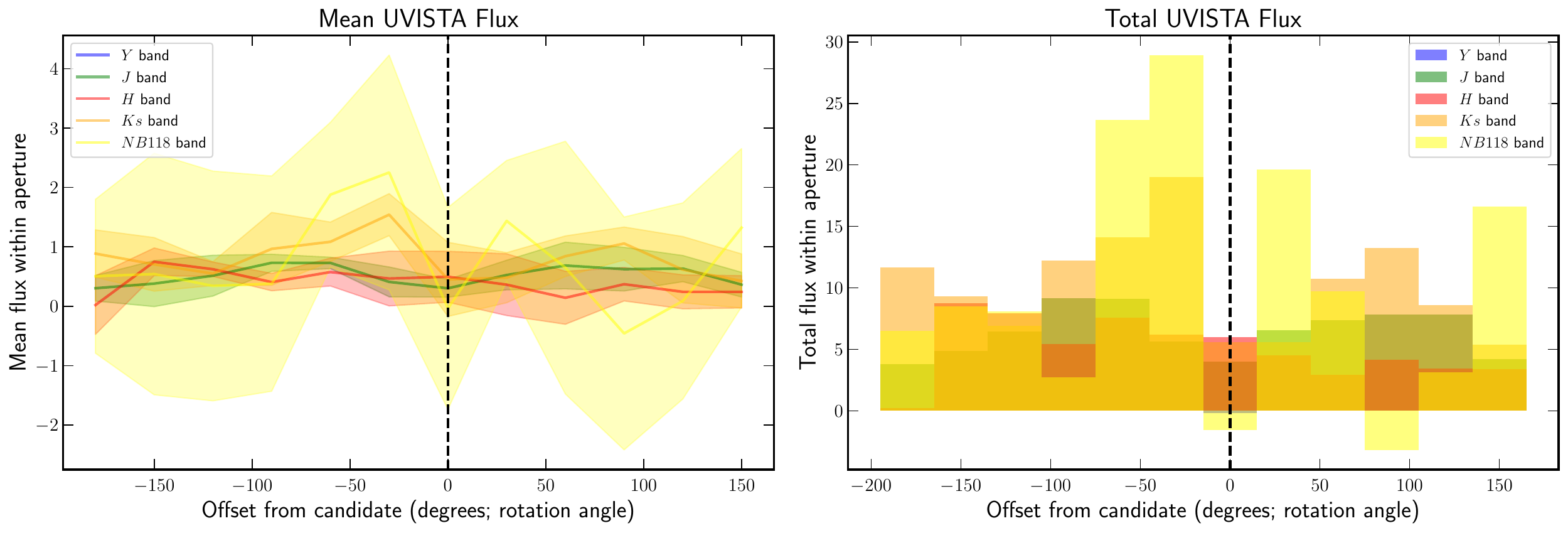}
    \caption{Flux from UltraVISTA imaging in the apertures from Figure \ref{fig:aperture_plot}, with $0^\circ$ corresponding to the location of HZTDE-1. The left figure is the mean flux within each aperture for each filter, and the right figure is the total flux within each aperture. Neither has an increase in flux at the location of HZTDE-1 in any filter.}
    \label{fig:aperture_flux}
\end{figure}

\newpage
\bibliography{highztde}{}
\bibliographystyle{aasjournal}

\end{document}